\documentclass[sigconf,authorversion]{acmart}

\clearpage{}

\usepackage[utf8]{inputenc}
\usepackage{textgreek}
\usepackage{fixltx2e} 
\usepackage{ifpdf}
\usepackage{xspace}

\newif\ifhtml
\ifpdf
\htmlfalse
\else
\htmltrue
\fi

\newcommand{\ie}{i.e.\xspace}
\newcommand{\eg}{e.g.\xspace}

\usepackage{letltxmacro}
\LetLtxMacro{\OrgCitep}{\citep}
\renewcommand{\citep}[1]{\,\OrgCitep{#1}}

\newcommand{\code}[1]{\texttt{#1}}
\usepackage{listings}
\usepackage{ifthen}
\newcommand{\citeurl}[5]{
#1\footnote{\emph{#2}
          \ifthenelse{\equal{#3}{}}
                     {}
                     {, #3}
          \ifthenelse{\equal{#4}{}}
                     {}
                     {, access date: #4}
          , \url{#5}}}

\makeatletter
\@ifpackageloaded{tex4ht}
  {}
  {\usepackage{tikz}
\tikzstyle{every picture}+=[font=\sffamily]}
\makeatother

\definecolor{fgcolor}{rgb}{0.345, 0.345, 0.345}
\definecolor{messagecolor}{rgb}{0, 0, 0}
\definecolor{errorcolor}{rgb}{1, 0, 0}

\newenvironment{knitrout}{}{}

\usepackage{framed}

\usepackage{alltt}
 
\usepackage[nameinlink]{cleveref}
\Crefformat{lstlisting}{#2Lst.\,#1#3}
\crefformat{lstlisting}{#2lst.\,#1#3}

\clearpage{}

\clearpage{}

\def\AwfyBaseline{
\begin{knitrout}
\definecolor{shadecolor}{rgb}{0.969, 0.969, 0.969}
\begin{tikzpicture}[x=1pt,y=1pt]
\definecolor{fillColor}{RGB}{255,255,255}
\path[use as bounding box,fill=fillColor,fill opacity=0.00] (0,0) rectangle ( 79.50,158.99);
\begin{scope}
\path[clip] ( 31.10, 30.99) rectangle ( 79.50,158.99);
\definecolor{drawColor}{gray}{0.80}

\path[draw=drawColor,line width= 0.6pt,dash pattern=on 4pt off 4pt ,line join=round] ( 31.10, 36.81) -- ( 79.50, 36.81);

\path[draw=drawColor,line width= 0.6pt,dash pattern=on 4pt off 4pt ,line join=round] ( 31.10, 55.06) -- ( 79.50, 55.06);

\path[draw=drawColor,line width= 0.6pt,dash pattern=on 4pt off 4pt ,line join=round] ( 31.10, 73.32) -- ( 79.50, 73.32);
\definecolor{drawColor}{gray}{0.20}

\path[draw=drawColor,line width= 0.6pt,line join=round] ( 40.17, 36.81) -- ( 40.17, 36.81);

\path[draw=drawColor,line width= 0.6pt,line join=round] ( 40.17, 36.81) -- ( 40.17, 36.81);
\definecolor{fillColor}{RGB}{255,255,255}

\path[draw=drawColor,line width= 0.6pt,line join=round,line cap=round,fill=fillColor] ( 34.50, 36.81) --
	( 34.50, 36.81) --
	( 45.84, 36.81) --
	( 45.84, 36.81) --
	( 34.50, 36.81) --
	cycle;

\path[draw=drawColor,line width= 1.1pt,line join=round] ( 34.50, 36.81) -- ( 45.84, 36.81);
\definecolor{fillColor}{gray}{0.20}

\path[draw=drawColor,line width= 0.4pt,line join=round,line cap=round,fill=fillColor] ( 55.30, 81.70) circle (  0.89);

\path[draw=drawColor,line width= 0.4pt,line join=round,line cap=round,fill=fillColor] ( 55.30, 79.25) circle (  0.89);

\path[draw=drawColor,line width= 0.4pt,line join=round,line cap=round,fill=fillColor] ( 55.30,153.18) circle (  0.89);

\path[draw=drawColor,line width= 0.4pt,line join=round,line cap=round,fill=fillColor] ( 55.30, 37.89) circle (  0.89);

\path[draw=drawColor,line width= 0.6pt,line join=round] ( 55.30, 63.87) -- ( 55.30, 65.13);

\path[draw=drawColor,line width= 0.6pt,line join=round] ( 55.30, 54.40) -- ( 55.30, 48.17);
\definecolor{fillColor}{RGB}{255,255,255}

\path[draw=drawColor,line width= 0.6pt,line join=round,line cap=round,fill=fillColor] ( 49.63, 63.87) --
	( 49.63, 54.40) --
	( 60.97, 54.40) --
	( 60.97, 63.87) --
	( 49.63, 63.87) --
	cycle;

\path[draw=drawColor,line width= 1.1pt,line join=round] ( 49.63, 57.99) -- ( 60.97, 57.99);

\path[draw=drawColor,line width= 0.6pt,line join=round] ( 70.42, 69.42) -- ( 70.42, 94.72);

\path[draw=drawColor,line width= 0.6pt,line join=round] ( 70.42, 47.71) -- ( 70.42, 42.07);

\path[draw=drawColor,line width= 0.6pt,line join=round,line cap=round,fill=fillColor] ( 64.75, 69.42) --
	( 64.75, 47.71) --
	( 76.09, 47.71) --
	( 76.09, 69.42) --
	( 64.75, 69.42) --
	cycle;

\path[draw=drawColor,line width= 1.1pt,line join=round] ( 64.75, 59.95) -- ( 76.09, 59.95);
\end{scope}
\begin{scope}
\path[clip] (  0.00,  0.00) rectangle ( 79.50,158.99);
\definecolor{drawColor}{RGB}{190,190,190}

\path[draw=drawColor,line width= 0.6pt,line join=round] ( 31.10, 30.99) --
	( 31.10,158.99);
\end{scope}
\begin{scope}
\path[clip] (  0.00,  0.00) rectangle ( 79.50,158.99);
\definecolor{drawColor}{gray}{0.30}

\node[text=drawColor,anchor=base east,inner sep=0pt, outer sep=0pt, scale=  0.80] at ( 26.15, 34.05) {1};

\node[text=drawColor,anchor=base east,inner sep=0pt, outer sep=0pt, scale=  0.80] at ( 26.15, 52.31) {2};

\node[text=drawColor,anchor=base east,inner sep=0pt, outer sep=0pt, scale=  0.80] at ( 26.15, 70.56) {3};

\node[text=drawColor,anchor=base east,inner sep=0pt, outer sep=0pt, scale=  0.80] at ( 26.15, 88.82) {4};

\node[text=drawColor,anchor=base east,inner sep=0pt, outer sep=0pt, scale=  0.80] at ( 26.15,107.07) {5};

\node[text=drawColor,anchor=base east,inner sep=0pt, outer sep=0pt, scale=  0.80] at ( 26.15,125.33) {6};

\node[text=drawColor,anchor=base east,inner sep=0pt, outer sep=0pt, scale=  0.80] at ( 26.15,143.58) {7};
\end{scope}
\begin{scope}
\path[clip] (  0.00,  0.00) rectangle ( 79.50,158.99);
\definecolor{drawColor}{gray}{0.20}

\path[draw=drawColor,line width= 0.6pt,line join=round] ( 28.35, 36.81) --
	( 31.10, 36.81);

\path[draw=drawColor,line width= 0.6pt,line join=round] ( 28.35, 55.06) --
	( 31.10, 55.06);

\path[draw=drawColor,line width= 0.6pt,line join=round] ( 28.35, 73.32) --
	( 31.10, 73.32);

\path[draw=drawColor,line width= 0.6pt,line join=round] ( 28.35, 91.57) --
	( 31.10, 91.57);

\path[draw=drawColor,line width= 0.6pt,line join=round] ( 28.35,109.83) --
	( 31.10,109.83);

\path[draw=drawColor,line width= 0.6pt,line join=round] ( 28.35,128.08) --
	( 31.10,128.08);

\path[draw=drawColor,line width= 0.6pt,line join=round] ( 28.35,146.34) --
	( 31.10,146.34);
\end{scope}
\begin{scope}
\path[clip] (  0.00,  0.00) rectangle ( 79.50,158.99);
\definecolor{drawColor}{RGB}{190,190,190}

\path[draw=drawColor,line width= 0.6pt,line join=round] ( 31.10, 30.99) --
	( 79.50, 30.99);
\end{scope}
\begin{scope}
\path[clip] (  0.00,  0.00) rectangle ( 79.50,158.99);
\definecolor{drawColor}{gray}{0.20}

\path[draw=drawColor,line width= 0.6pt,line join=round] ( 40.17, 28.24) --
	( 40.17, 30.99);

\path[draw=drawColor,line width= 0.6pt,line join=round] ( 55.30, 28.24) --
	( 55.30, 30.99);

\path[draw=drawColor,line width= 0.6pt,line join=round] ( 70.42, 28.24) --
	( 70.42, 30.99);
\end{scope}
\begin{scope}
\path[clip] (  0.00,  0.00) rectangle ( 79.50,158.99);
\definecolor{drawColor}{gray}{0.30}

\node[text=drawColor,rotate= 90.00,anchor=base east,inner sep=0pt, outer sep=0pt, scale=  0.80] at ( 42.93, 26.04) {Java};

\node[text=drawColor,rotate= 90.00,anchor=base east,inner sep=0pt, outer sep=0pt, scale=  0.80] at ( 58.05, 26.04) {Node.js};

\node[text=drawColor,rotate= 90.00,anchor=base east,inner sep=0pt, outer sep=0pt, scale=  0.80] at ( 73.18, 26.04) {SOMns};
\end{scope}
\begin{scope}
\path[clip] (  0.00,  0.00) rectangle ( 79.50,158.99);
\definecolor{drawColor}{RGB}{0,0,0}

\node[text=drawColor,rotate= 90.00,anchor=base,inner sep=0pt, outer sep=0pt, scale=  0.80] at (  5.51, 94.99) {Runtime Factor};

\node[text=drawColor,rotate= 90.00,anchor=base,inner sep=0pt, outer sep=0pt, scale=  0.80] at ( 14.15, 94.99) {normalized to Java (lower is better)};
\end{scope}
\end{tikzpicture}
 
\end{knitrout}
}
\clearpage{}
\clearpage{}

\def\SavinaBaseline{
\begin{knitrout}
\definecolor{shadecolor}{rgb}{0.969, 0.969, 0.969}

\end{knitrout}
}

\newcommand{\TraceOverheadSidGMeanP}{10\%\xspace}
\newcommand{\TraceOverheadSidMinP}{0\%\xspace}
\newcommand{\TraceOverheadSidMaxP}{20\%\xspace}

\newcommand{\TraceOverheadFidGMeanP}{9\%\xspace}
\newcommand{\TraceOverheadFidMinP}{0\%\xspace}
\newcommand{\TraceOverheadFidMaxP}{18\%\xspace}

\def\SavinaTraceDataTable{
\begin{tabular}{lc c }
\hline
 & \multicolumn{2}{r}{Harmonic Mean MB/s} \\ 
  & \multicolumn{1}{r}{Small Ids} & \multicolumn{1}{r}{Full Ids} \\ 
\hline
BankTransaction  & \multicolumn{1}{r}{19.11} & \multicolumn{1}{r}{23.87} \\
BigContention  & \multicolumn{1}{r}{10.38} & \multicolumn{1}{r}{12.14} \\
Chameneos  & \multicolumn{1}{r}{52.02} & \multicolumn{1}{r}{53.46} \\
CigaretteSmokers  & \multicolumn{1}{r}{30.72} & \multicolumn{1}{r}{31.50} \\
ConcurrentDictionary  & \multicolumn{1}{r}{0.02} & \multicolumn{1}{r}{0.03} \\
ConcurrentSortedLinkedList  & \multicolumn{1}{r}{0.00} & \multicolumn{1}{r}{4.38} \\
Counting  & \multicolumn{1}{r}{58.84} & \multicolumn{1}{r}{108.58} \\
ForkJoinActorCreation  & \multicolumn{1}{r}{18.56} & \multicolumn{1}{r}{34.23} \\
ForkJoinThroughput  & \multicolumn{1}{r}{7.96} & \multicolumn{1}{r}{25.44} \\
LogisticMapSeries  & \multicolumn{1}{r}{30.64} & \multicolumn{1}{r}{49.00} \\
Philosophers  & \multicolumn{1}{r}{44.52} & \multicolumn{1}{r}{71.26} \\
PingPong  & \multicolumn{1}{r}{39.22} & \multicolumn{1}{r}{68.18} \\
ProducerConsumerBoundedBuffer  & \multicolumn{1}{r}{5.29} & \multicolumn{1}{r}{0.02} \\
RadixSort  & \multicolumn{1}{r}{36.29} & \multicolumn{1}{r}{66.44} \\
SleepingBarber  & \multicolumn{1}{r}{55.88} & \multicolumn{1}{r}{70.42} \\
ThreadRing  & \multicolumn{1}{r}{50.00} & \multicolumn{1}{r}{76.91} \\
TrapezoidalApproximation  & \multicolumn{1}{r}{3.17} & \multicolumn{1}{r}{3.21} \\
UnbalancedCobwebbedTree  & \multicolumn{1}{r}{19.60} & \multicolumn{1}{r}{20.02} \\
\hline 
\end{tabular}

}

\newcommand{\TraceRateFidMaxMBS}
  {109\,MB/s\xspace}
\newcommand{\TraceRateSidMaxMBS}
  {59\,MB/s\xspace}
\clearpage{}
\clearpage{}

\def\AcmeTracing{
\begin{knitrout}
\definecolor{shadecolor}{rgb}{0.969, 0.969, 0.969}
\begin{tikzpicture}[x=1pt,y=1pt]
\definecolor{fillColor}{RGB}{255,255,255}
\path[use as bounding box,fill=fillColor,fill opacity=0.00] (0,0) rectangle (231.26,144.54);
\begin{scope}
\path[clip] ( 78.23, 35.82) rectangle (231.26,144.54);
\definecolor{drawColor}{gray}{0.20}

\path[draw=drawColor,line width= 0.6pt,line join=round] (163.31, 43.77) -- (177.94, 43.77);

\path[draw=drawColor,line width= 0.6pt,line join=round] (134.06, 43.77) -- (119.43, 43.77);
\definecolor{fillColor}{RGB}{255,255,255}

\path[draw=drawColor,line width= 0.6pt,line join=round,line cap=round,fill=fillColor] (163.31, 38.80) --
	(134.06, 38.80) --
	(134.06, 48.75) --
	(163.31, 48.75) --
	(163.31, 38.80) --
	cycle;

\path[draw=drawColor,line width= 1.1pt,line join=round] (148.68, 38.80) -- (148.68, 48.75);

\path[draw=drawColor,line width= 0.6pt,line join=round] (171.02, 57.03) -- (177.94, 57.03);

\path[draw=drawColor,line width= 0.6pt,line join=round] (157.18, 57.03) -- (150.26, 57.03);

\path[draw=drawColor,line width= 0.6pt,line join=round,line cap=round,fill=fillColor] (171.02, 52.06) --
	(157.18, 52.06) --
	(157.18, 62.00) --
	(171.02, 62.00) --
	(171.02, 52.06) --
	cycle;

\path[draw=drawColor,line width= 1.1pt,line join=round] (164.10, 52.06) -- (164.10, 62.00);

\path[draw=drawColor,line width= 0.6pt,line join=round] (175.02, 70.29) -- (177.94, 70.29);

\path[draw=drawColor,line width= 0.6pt,line join=round] (169.19, 70.29) -- (166.28, 70.29);

\path[draw=drawColor,line width= 0.6pt,line join=round,line cap=round,fill=fillColor] (175.02, 65.32) --
	(169.19, 65.32) --
	(169.19, 75.26) --
	(175.02, 75.26) --
	(175.02, 65.32) --
	cycle;

\path[draw=drawColor,line width= 1.1pt,line join=round] (172.11, 65.32) -- (172.11, 75.26);

\path[draw=drawColor,line width= 0.6pt,line join=round] (156.13, 83.55) -- (177.94, 83.55);

\path[draw=drawColor,line width= 0.6pt,line join=round] (112.53, 83.55) -- ( 90.73, 83.55);

\path[draw=drawColor,line width= 0.6pt,line join=round,line cap=round,fill=fillColor] (156.13, 78.58) --
	(112.53, 78.58) --
	(112.53, 88.52) --
	(156.13, 88.52) --
	(156.13, 78.58) --
	cycle;

\path[draw=drawColor,line width= 1.1pt,line join=round] (134.33, 78.58) -- (134.33, 88.52);

\path[draw=drawColor,line width= 0.6pt,line join=round] (209.05, 96.81) -- (219.42, 96.81);

\path[draw=drawColor,line width= 0.6pt,line join=round] (188.31, 96.81) -- (177.94, 96.81);

\path[draw=drawColor,line width= 0.6pt,line join=round,line cap=round,fill=fillColor] (209.05, 91.84) --
	(188.31, 91.84) --
	(188.31,101.78) --
	(209.05,101.78) --
	(209.05, 91.84) --
	cycle;

\path[draw=drawColor,line width= 1.1pt,line join=round] (198.68, 91.84) -- (198.68,101.78);

\path[draw=drawColor,line width= 0.6pt,line join=round] (169.24,110.07) -- (177.94,110.07);

\path[draw=drawColor,line width= 0.6pt,line join=round] (151.86,110.07) -- (143.16,110.07);

\path[draw=drawColor,line width= 0.6pt,line join=round,line cap=round,fill=fillColor] (169.24,105.10) --
	(151.86,105.10) --
	(151.86,115.04) --
	(169.24,115.04) --
	(169.24,105.10) --
	cycle;

\path[draw=drawColor,line width= 1.1pt,line join=round] (160.55,105.10) -- (160.55,115.04);

\path[draw=drawColor,line width= 0.6pt,line join=round] (180.40,123.33) -- (181.23,123.33);

\path[draw=drawColor,line width= 0.6pt,line join=round] (178.76,123.33) -- (177.94,123.33);

\path[draw=drawColor,line width= 0.6pt,line join=round,line cap=round,fill=fillColor] (180.40,118.35) --
	(178.76,118.35) --
	(178.76,128.30) --
	(180.40,128.30) --
	(180.40,118.35) --
	cycle;

\path[draw=drawColor,line width= 1.1pt,line join=round] (179.58,118.35) -- (179.58,128.30);

\path[draw=drawColor,line width= 0.6pt,line join=round] (168.79,136.58) -- (177.94,136.58);

\path[draw=drawColor,line width= 0.6pt,line join=round] (150.51,136.58) -- (141.37,136.58);

\path[draw=drawColor,line width= 0.6pt,line join=round,line cap=round,fill=fillColor] (168.79,131.61) --
	(150.51,131.61) --
	(150.51,141.56) --
	(168.79,141.56) --
	(168.79,131.61) --
	cycle;

\path[draw=drawColor,line width= 1.1pt,line join=round] (159.65,131.61) -- (159.65,141.56);
\definecolor{drawColor}{gray}{0.80}

\path[draw=drawColor,line width= 0.6pt,dash pattern=on 4pt off 4pt ,line join=round] (177.94, 35.82) -- (177.94,144.54);
\end{scope}
\begin{scope}
\path[clip] (  0.00,  0.00) rectangle (231.26,144.54);
\definecolor{drawColor}{RGB}{190,190,190}

\path[draw=drawColor,line width= 0.6pt,line join=round] ( 78.23, 35.82) --
	( 78.23,144.54);
\end{scope}
\begin{scope}
\path[clip] (  0.00,  0.00) rectangle (231.26,144.54);
\definecolor{drawColor}{gray}{0.30}

\node[text=drawColor,anchor=base east,inner sep=0pt, outer sep=0pt, scale=  0.80] at ( 73.28, 41.02) {BookFlight};

\node[text=drawColor,anchor=base east,inner sep=0pt, outer sep=0pt, scale=  0.80] at ( 73.28, 54.28) {Cancel Booking};

\node[text=drawColor,anchor=base east,inner sep=0pt, outer sep=0pt, scale=  0.80] at ( 73.28, 67.54) {List Bookings};

\node[text=drawColor,anchor=base east,inner sep=0pt, outer sep=0pt, scale=  0.80] at ( 73.28, 80.79) {Login};

\node[text=drawColor,anchor=base east,inner sep=0pt, outer sep=0pt, scale=  0.80] at ( 73.28, 94.05) {Query Flight};

\node[text=drawColor,anchor=base east,inner sep=0pt, outer sep=0pt, scale=  0.80] at ( 73.28,107.31) {Update Customer};

\node[text=drawColor,anchor=base east,inner sep=0pt, outer sep=0pt, scale=  0.80] at ( 73.28,120.57) {View Profile};

\node[text=drawColor,anchor=base east,inner sep=0pt, outer sep=0pt, scale=  0.80] at ( 73.28,133.83) {Logout};
\end{scope}
\begin{scope}
\path[clip] (  0.00,  0.00) rectangle (231.26,144.54);
\definecolor{drawColor}{gray}{0.20}

\path[draw=drawColor,line width= 0.6pt,line join=round] ( 75.48, 43.77) --
	( 78.23, 43.77);

\path[draw=drawColor,line width= 0.6pt,line join=round] ( 75.48, 57.03) --
	( 78.23, 57.03);

\path[draw=drawColor,line width= 0.6pt,line join=round] ( 75.48, 70.29) --
	( 78.23, 70.29);

\path[draw=drawColor,line width= 0.6pt,line join=round] ( 75.48, 83.55) --
	( 78.23, 83.55);

\path[draw=drawColor,line width= 0.6pt,line join=round] ( 75.48, 96.81) --
	( 78.23, 96.81);

\path[draw=drawColor,line width= 0.6pt,line join=round] ( 75.48,110.07) --
	( 78.23,110.07);

\path[draw=drawColor,line width= 0.6pt,line join=round] ( 75.48,123.33) --
	( 78.23,123.33);

\path[draw=drawColor,line width= 0.6pt,line join=round] ( 75.48,136.58) --
	( 78.23,136.58);
\end{scope}
\begin{scope}
\path[clip] (  0.00,  0.00) rectangle (231.26,144.54);
\definecolor{drawColor}{RGB}{190,190,190}

\path[draw=drawColor,line width= 0.6pt,line join=round] ( 78.23, 35.82) --
	(231.26, 35.82);
\end{scope}
\begin{scope}
\path[clip] (  0.00,  0.00) rectangle (231.26,144.54);
\definecolor{drawColor}{gray}{0.20}

\path[draw=drawColor,line width= 0.6pt,line join=round] ( 85.19, 33.07) --
	( 85.19, 35.82);

\path[draw=drawColor,line width= 0.6pt,line join=round] (116.10, 33.07) --
	(116.10, 35.82);

\path[draw=drawColor,line width= 0.6pt,line join=round] (147.02, 33.07) --
	(147.02, 35.82);

\path[draw=drawColor,line width= 0.6pt,line join=round] (177.94, 33.07) --
	(177.94, 35.82);

\path[draw=drawColor,line width= 0.6pt,line join=round] (208.85, 33.07) --
	(208.85, 35.82);
\end{scope}
\begin{scope}
\path[clip] (  0.00,  0.00) rectangle (231.26,144.54);
\definecolor{drawColor}{gray}{0.30}

\node[text=drawColor,rotate= 90.00,anchor=base east,inner sep=0pt, outer sep=0pt, scale=  0.80] at ( 87.94, 30.87) {0.97};

\node[text=drawColor,rotate= 90.00,anchor=base east,inner sep=0pt, outer sep=0pt, scale=  0.80] at (118.86, 30.87) {0.98};

\node[text=drawColor,rotate= 90.00,anchor=base east,inner sep=0pt, outer sep=0pt, scale=  0.80] at (149.77, 30.87) {0.99};

\node[text=drawColor,rotate= 90.00,anchor=base east,inner sep=0pt, outer sep=0pt, scale=  0.80] at (180.69, 30.87) {1.00};

\node[text=drawColor,rotate= 90.00,anchor=base east,inner sep=0pt, outer sep=0pt, scale=  0.80] at (211.61, 30.87) {1.01};
\end{scope}
\begin{scope}
\path[clip] (  0.00,  0.00) rectangle (231.26,144.54);
\definecolor{drawColor}{RGB}{0,0,0}

\node[text=drawColor,anchor=base,inner sep=0pt, outer sep=0pt, scale=  0.80] at (154.75,  9.89) {Latency Factor};

\node[text=drawColor,anchor=base,inner sep=0pt, outer sep=0pt, scale=  0.80] at (154.75,  1.25) {normalized to untraced (lower is better)};
\end{scope}
\end{tikzpicture}
 
\end{knitrout}
}

\newcommand{\AcmeLatencyTracedGMeanP}{
  -1\%\xspace}

\newcommand{\AcmeLatencyTracedMinP}{
  -3\%\xspace}

\newcommand{\AcmeLatencyTracedMaxP}{
  1\%\xspace}

\newcommand{\AcmeQueryFP}{
  46.73\%\xspace}

\clearpage{}
\clearpage{}

\newcommand{\RosaGMeanX}{1.56x\xspace}
\newcommand{\RosaMinX}{0.93x\xspace}
\newcommand{\RosaMaxX}{2.08x\xspace}
\clearpage{}

\def\SOMns{SOM{\sc ns}\xspace}
\def\Kompos{Kómpos\xspace}

\copyrightyear{2018} 
\acmYear{2018} 
\setcopyright{acmlicensed}
\acmConference[ManLang'18]{15th International Conference on Managed Languages \& Runtimes}{September 12--14, 2018}{Linz, Austria}
\acmBooktitle{15th International Conference on Managed Languages \& Runtimes (ManLang'18), September 12--14, 2018, Linz, Austria}
\acmPrice{15.00}
\acmDOI{10.1145/3237009.3237015}
\acmISBN{978-1-4503-6424-9/18/09}

\usepackage{booktabs}   
\usepackage{subcaption}

\begin{document}

\title{Efficient and Deterministic Record \& Replay for Actor Languages}

\author{Dominik Aumayr}
\affiliation{
  \institution{Johannes Kepler University}            
  \streetaddress{}
  \city{Linz}
  \country{Austria}                    
}
\email{dominik.aumayr@jku.at}

\author{Stefan Marr}
\affiliation{
	\institution{University of Kent}            
	\city{Canterbury}
	\country{United Kingdom}                    
}
\email{s.marr@kent.ac.uk}          

\author{Cl\'ement B\'era}
\affiliation{
	\institution{Vrije Universiteit Brussel}            
	\city{Brussel}
	\country{Belgium}                    
}
\email{clement.bera@vub.be}          

\author{Elisa Gonzalez Boix}
\affiliation{
	\institution{Vrije Universiteit Brussel}            
	\city{Brussel}
	\country{Belgium}                    
}
\email{egonzale@vub.be}          

\author{Hanspeter M\"{o}ssenb\"{o}ck}
\affiliation{
  \institution{Johannes Kepler University}           
  \city{Linz}
  \country{Austria}                   
}
\email{hanspeter.moessenboeck@jku.at}

\renewcommand{\shortauthors}{D. Aumayr et al.}

\begin{abstract}
	
With the ubiquity of parallel commodity hardware,
developers turn to high-level concurrency models such as the actor model to lower the complexity of concurrent software.
However, debugging concurrent software is hard,
especially for concurrency models with a limited set of supporting tools.
Such tools often deal only with the underlying threads and locks, 
which obscures the view on \eg actors and messages
and thereby introduces additional complexity.

To improve on this situation,
we present a low-overhead record \& replay approach for actor languages.
It allows one to debug concurrency issues deterministically based on a previously recorded trace.
Our evaluation shows that the average run-time overhead for tracing on benchmarks from the Savina suite is \TraceOverheadSidGMeanP (min. \TraceOverheadSidMinP, max. \TraceOverheadSidMaxP). 
For Acme-Air, a modern web application, we see a maximum increase of\AcmeLatencyTracedMaxP in latency for HTTP requests and about 1.4\,MB/s of trace data.
These results are a first step towards deterministic replay debugging of actor systems in production.

\end{abstract}

\ccsdesc[500]{Computing methodologies~Concurrent programming languages}
\ccsdesc[300]{Software and its engineering~Software testing and debugging}

\keywords{Concurrency, Debugging, Determinism, Actors, Tracing, Replay}

\maketitle

\section{Introduction}
Debugging concurrent systems is hard, 
because they can be non-deterministic, 
and so can be the bugs one tries to fix. 
The main challenge with these so called \emph{Heisenbugs}\citep{gray1986computers}, is 
that they may manifest rarely and may disappear during debugging,
which makes them hard to reproduce and to fix.

\citet{McDowell:1989:DCP} 
distinguish two broad categories of debuggers for finding and fixing bugs:
traditional breakpoint-based debuggers and event-based debuggers.
Event-based debuggers see a program execution as a series of events
and abstract away implementation details.
Commonly such event traces are used for post-mortem analyses.
However, they can also be used to reproduce program execution,
which is known as \emph{record \& replay}.
With record \& replay it is possible to repeat a recorded execution arbitrarily often.
Therefore, once a program execution with a manifested bug was recorded,
the bug can be reproduced reliably.
This makes such bugs easier to locate even though many executions may need to be recorded to capture the bug.

Record \& replay has been investigated in the past\citep{Chen:2015:DRS}
for thread-based programs or message-passing systems, at least since the 1980s\citep{Curtis:1982:bugnet}.
However, debugging support for high-level concurrency models such as the actor model
has not yet received as much attention\citep{TorresLopez:2016:TAD}. 
As a result, there is a lack of appropriate tools,
which poses a maintenance challenge for complex systems.
This is problematic because
popular implementations of the actor model, such as
Akka\citeurl{}{Akka website}{}{}{https://akka.io/}, 
Pony\citep{Clebsch:2015:DCS:2824815.2824816}, 
Erlang\citep{Erlang}, 
Elixir\citep{Thomas:2014:ELIXIR}, 
Orleans\citep{Bykov:2011:OCC:2038916.2038932}, 
and Node.js\citeurl{}{Node.js website}{}{}{https://nodejs.org/},  
are used to build increasingly
complex server applications.

Debugging support for the actor model so far focused either on
breakpoint-based debuggers with support for
actor-specific inspection, stepping operations,
breakpoints, asynchronous stack traces, and visualizations\citep{Boix:2011:REMED,Marr:2017:Kompos},
or it focused on postmortem debugging, \eg Causeway\citep{stanley2009causeway},
where a program's execution is analyzed after it crashed.
While specialized debuggers provide us with the ability to inspect the execution of actor programs,
they do not tackle non-determinism.
However, to the best of our knowledge,
existing record \& replay approaches for actor-based systems
focus either on single event loop environments\citep{Burg:2013:IRW,barr:2016:time}
or have not yet considered the performance requirements for
server applications\citep{shibanai2017actoverse}.

In this paper, we present an \emph{efficient} approach for recording \& replaying concurrent actor-based systems.
By tracing and reproducing the ordering of messages,
recording of application data can be limited to I/O operations. 
To minimize the run-time overhead,
we determine a small set of events needed to replay actor applications.
We prototype our approach on an implementation of
communicating event loop actors\citep{Miller:2005:CSP} in \SOMns. 
\SOMns is an implementation of Newspeak\citep{Bracha:2010:ECOOP} on top of the Truffle framework and the Graal just-in-time compiler\citep{Wurthinger:2017:PPE}.
Furthermore, we provide support for recording additional detailed information during replay executions,
which can be used in the \Kompos debugger\citep{Marr:2017:Kompos} for visualizations or post-mortem analyses. 

We evaluate our approach
with \SOMns.
Using the Savina micro-benchmark suite\citep{Imam:2014:SAB}, 
we measure 
the tracing run-time overhead
and the trace growth rate 
for each benchmark.
On the Acme-Air web application\citep{Ueda:2016:AcmeAir}, 
we measure the latency with and without tracing,
and the total trace size recorded.

The contributions of our approach are:
\begin{enumerate}
	\item Deterministic replay of actor applications using high-level messaging abstractions,
	\item Capture of non-deterministic data to deal with external inputs,
	\item Scalability to a high number of actors and messages.
\end{enumerate}

\section{Towards Efficient Deterministic Replay for Actor Languages}
\label{sec:towards}

In deterministic programs, the result of an execution depends only on its input.
Thus, reproducing an execution is straightforward, 
provided the environment is the same.
In practice, 
it is often necessary to debug a program multiple times before the root cause of a bug is found. 
This approach to debugging is called \emph{cyclic debugging}\citep{McDowell:1989:DCP}. 
As convenient as cyclic debugging is, 
it requires bugs to be reproducible reliably. 
This makes it unsuitable for non-deterministic programs, 
where the occurrence of a bug may depend on a rare scheduling of messages.

As mentioned before, record \& replay\citep{Chen:2015:DRS}
enables deterministic re-execution of a previously recorded program execution,
and thereby enables cyclic debugging also for non-deterministic programs. 
During the initial execution, such approaches record a program trace,
which is then used during replay to guide the execution and reproduce,
for instance, input from external sources and scheduling decisions,
and thereby eliminate all non-determinism.

Record \& replay for parallel and concurrent programs has been studied before,
but a majority of the previous work focused on shared memory concurrency
and MPI-like message passing\citep{Chen:2015:DRS}.
Recent work focused either on single event loops
or did not consider performance\citep{Burg:2013:IRW,barr:2016:time,shibanai2017actoverse}.
Thus, none of the approaches that we are aware of support
efficient deterministic record \& replay 
for modern actor-based applications.

The remainder of this section considers the practical requirements
for an efficient deterministic record \& replay system.
Furthermore, it provides the necessary background on actor-based concurrency
and considers the limitations of record \& replay systems.

\subsection{Practical Requirements for Record~\&~Replay}
\label{sec:requirements}

Since modern actor systems such as Akka, Pony, Erlang, Elixir, Orleans, and Node.js
are widely used for server applications,
we aim at making it practical to record the execution of such applications.
In such an environment, bugs might occur rarely
and could be triggered by specific user interactions only.
We assume that development happens on commodity hardware, so that
the issues can be reproduced and debugged on a developer's laptop or workstation.

Based on this scenario, we consider two main concerns.
First, the recording should have minimal run-time overhead to minimize the effect
on possible Heisenbugs.
Second, the amount of recorded data should be small enough
to fit either into memory or on a commodity storage.
For comparison,
\citet{barr:2016:time} reported a maximal tracing overhead of 2\% for their
single event loop Node.js system and 4-8 seconds of benchmark execution.
The produced trace data is less than 9 MB.
\citet{Burg:2013:IRW} report 1-3\% run-time overhead and in the worst case 65\%.
Their benchmarks execute for up to 26 seconds and produce up to 700 KB of traces.

To make our system practical, we aim to achieve a similarly small run-time overhead
while tracing multiple event loops.
However, in parallel actor applications,
we need to account for much higher degrees of non-determinism.
This means that the run-time overhead is likely larger.
Additionally, run-time overhead can scale
  with the tracing workload,
  for instance, 
  message intensive programs
  may have a higher overhead
  than computationally intensive ones.
Thus, our goal is:

\vspace{0.2cm}
\fbox{\parbox{0.9\linewidth}{\center{
\textbf{Goal 1}

\emph{The run-time overhead of tracing for server applications should be in the 0\% to 3\% range.
Worst-case run-time overhead, e.g. for message intensive programs, should be below 25\%.}}}}
\vspace{0.2cm}

Since we aim at supporting long-running actor-based server applications,
the reported trace sizes do not directly compare to our scenario.
Furthermore, they are based on single event loops, which have much lower event rates.
Since we assume that some bugs might be induced by user interactions,
we want to support executions of multiple minutes and perhaps up to half an hour.
Considering that contemporary laptops have about 500 GB of storage,
this would mean an execution should produce no more than about 250 MB/s of trace data.
Therefore, our second goal is:

\vspace{0.2cm}
\fbox{\parbox{0.9\linewidth}{\center{
\textbf{Goal 2}

\emph{Recording should produce well below 250 MB/s of trace data.}

 }}}

\subsection{Communicating Event Loop Actors}

This section provides the background on actor-based concurrency
to detail the challenges of designing an efficient record \& replay mechanism
for actor languages, and our contributions.

The actor model of concurrency was first proposed by \citet{ActorsFormalism}. 
By now, diverse variations have emerged\citep{DeKoster:2016:YAT}.
We focus on the communicating event loops (CEL) variant
pioneered by the language E\citep{Miller:2005:CSP}.
The CEL model exhibits all relevant characteristics of actor models
and combines event loops with high-level abstractions, like non-blocking promises,
which represent a challenge for deterministic replay, as we detail below.
This class of actor models has been later adopted by languages such as 
AmbientTalk\citep{VanCutsem:2012:AMA} and Newspeak\citep{Bracha:2010:ECOOP},
and also corresponds to the asynchronous programming model of JavaScript and Node.js\citep{Tilkov:2010:NodeJs}.

\begin{figure}[t!]
	\centering
	\includegraphics[width=\linewidth]{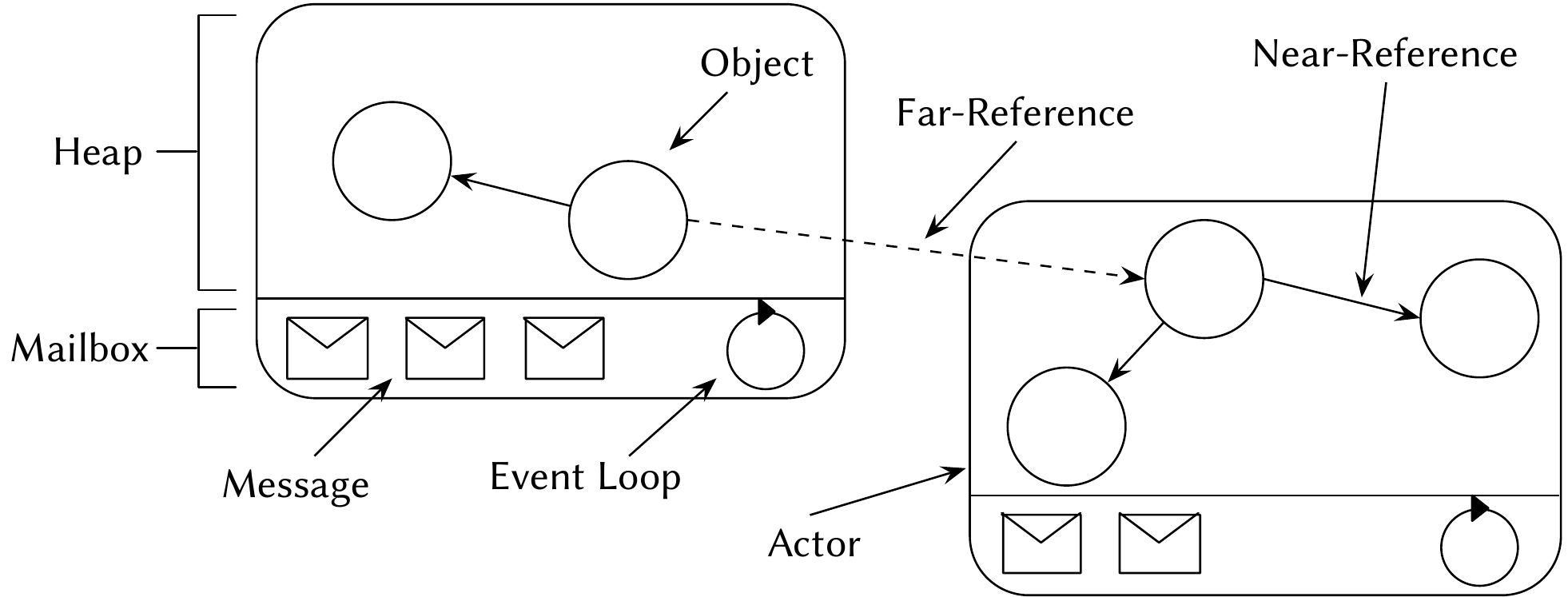} 
	\caption{Overview of CEL model. Each actor consists of a thread of execution (an event loop), a heap with regular objects, and a mailbox. An event loop processes incoming messages in a serial order from its mailbox. An actor can directly access and mutate objects it owns. All communication with objects owned by other actors happens asynchronously via far references. 
	}
	\label{fig:cel}
\end{figure}

The general structure of CEL is shown in \cref{fig:cel}.
Each actor is a container of objects isolated from the others, a mailbox, and an event loop.
The event loop processes messages from its mailbox one-by-one in order of arrival.
When a message is processed,
the receiver object is identified and the method corresponding to the message is invoked.
The processing of one message by an actor defines a \emph{turn}.
Since actors have isolated state and messages are handled atomically with respect to other messages,
the non-determinism of the system is restricted to the order in which messages are processed. 

To maintain the state of each actor isolated from the other actors, each actor only has direct access to the objects it owns. Communication with objects owned by other actors happens via \emph{far references}. Far references do not support synchronous method invocation nor direct access to fields of objects. Instead, they can only receive asynchronous message sends, which are forwarded to the mailbox of the actor owning the object. Objects passed as arguments in asynchronous message sends are parameter-passed either by far reference, or by (deep) copy.

An asynchronous message send immediately returns a \emph{promise} (also know as a future).
A promise is a placeholder object for the result that is to be computed.
Once the return value is computed,
it is accessible through the promise, which is then said to be \emph{resolved} with the value.
The promise itself is an object, which can receive asynchronous messages.
Those messages are accumulated within the promise and forwarded to the result value once it is available.

Other actor variants have different semantics for message reception
and whether they support (non-blocking) promises.
Note, however, that the queuing on non-blocking promises introduces additional non-determinism compared to other actor variants. 
Thus, they are the most challenging variant for deterministic replay.

\subsection{Record \& Replay for Actors}
\label{sub:rr}

As mentioned before, record \& replay has been investigated before\citep{Chen:2015:DRS}.
\citet{ronsse:2000:execution} categorizes such approaches
into content-based and ordering-based replay based on what type of data is recorded.
We now describe their characteristics and applicability to actor-based concurrency. 

\paragraph{Content-based Replay}
\label{sec:content-based-replay}

Content-based replay is based on recording the results of all operations that observe non-determinism, 
and returning the recorded results during replay. 
In the context of shared memory concurrency, 
this means that all reads from memory accessed by other threads need to be captured. A representative example of such an approach is BugNet\citep{Narayanasamy:2005:BCR:1080695.1069994}.

In the context of actor-based concurrency, it is necessary to record all kinds of events received by actors.
To the best of our knowledge, there exist only three approaches providing record \& replay for actor-based concurrency: Jardis\citep{barr:2016:time}, Dolos\citep{Burg:2013:IRW} and Actoverse\citep{shibanai2017actoverse}.
They can be categorized as content-based replay.
Actoverse provides record \& replay for Akka programs and records messages exchanged by actors including message contents. 
Dolos does record \& replay for JavaScript applications running in a browser, and Jardis for both the browser and Node.js.
Both Dolos and Jardis capture all non-deterministic interactions within a single event loop, \ie interactions with JavaScript/Node.js APIs.

\paragraph{Ordering-based replay}
Ordering-based replay (also known as control-based replay) focuses on the order in which non-de\-ter\-min\-istic events occur.
The key idea is that by reproducing the control-flow of an execution, 
the data is implicitly reproduced as well. 
This means that only data needed to reproduce the control-flow has to be recorded, producing smaller traces in the process.
An early implementation of ordering-based replay is Instant replay\citep{leblanc:1987:debugging}, 
which maintains version numbers for shared memory variables. 
However, ordering-based replay does not work when a program has non-deterministic inputs.
For such programs, ordering-based replay can be used for internal non-determinism,
combined with content-based replay for non-deterministic inputs.

In actor-based concurrency, 
since the non-determinism of the system is restricted to the order in which messages are processed, 
it is only necessary to reproduce the message processing order of an actor. 
Ordering-based replay has not been explored for actor-based concurrency,
but there is work for message passing interface (MPI) libraries.
MPL*\citep{Kergommeaux:1999:MER} is an ordering-based record \& replay debugger for MPI communication. 
MPL* records the sequence of message origins (senders). 
This is enough information to reproduce the ordering of messages for MPI communication,
since messages from the same source are race-free,
\ie, they arrive in the order they were sent in.
Another ordering-based record \& replay approach for MPI is Reconstruction of Lamport Timestamps (ROLT)\citep{Ronsse:ROLT}. 
Like MPL*, ROLT assumes messages from the same source being race-free.
It then uses Lamport clocks in all actors, and records when a clock update is larger than one time step. These ``jumps'' are caused by communication with actors that have a different clock value, which synchronizes the Lamport clocks. In replay, 
the sending of a message is delayed until all messages with smaller timestamps have been sent.

\subsection{Problem Statement}
\label{sub:problem}

The existing record \& replay approaches discussed above
leave three issues that need to be solved for actor-based concurrency. 

\paragraph{Issue 1: Deterministic replay of high-level messaging abstractions.}

Existing record \& replay approaches typically only record the sequence of messages
to reproduce the message order.
MPL* for example only records message senders,
while Actoverse records message contents as well.
Unfortunately, message sender and content are not enough to reproduce
the original message ordering in the presence of high-level messaging abstractions such as promises.

\begin{figure}[tb]
		\centering
		\includegraphics[width=0.98\linewidth]{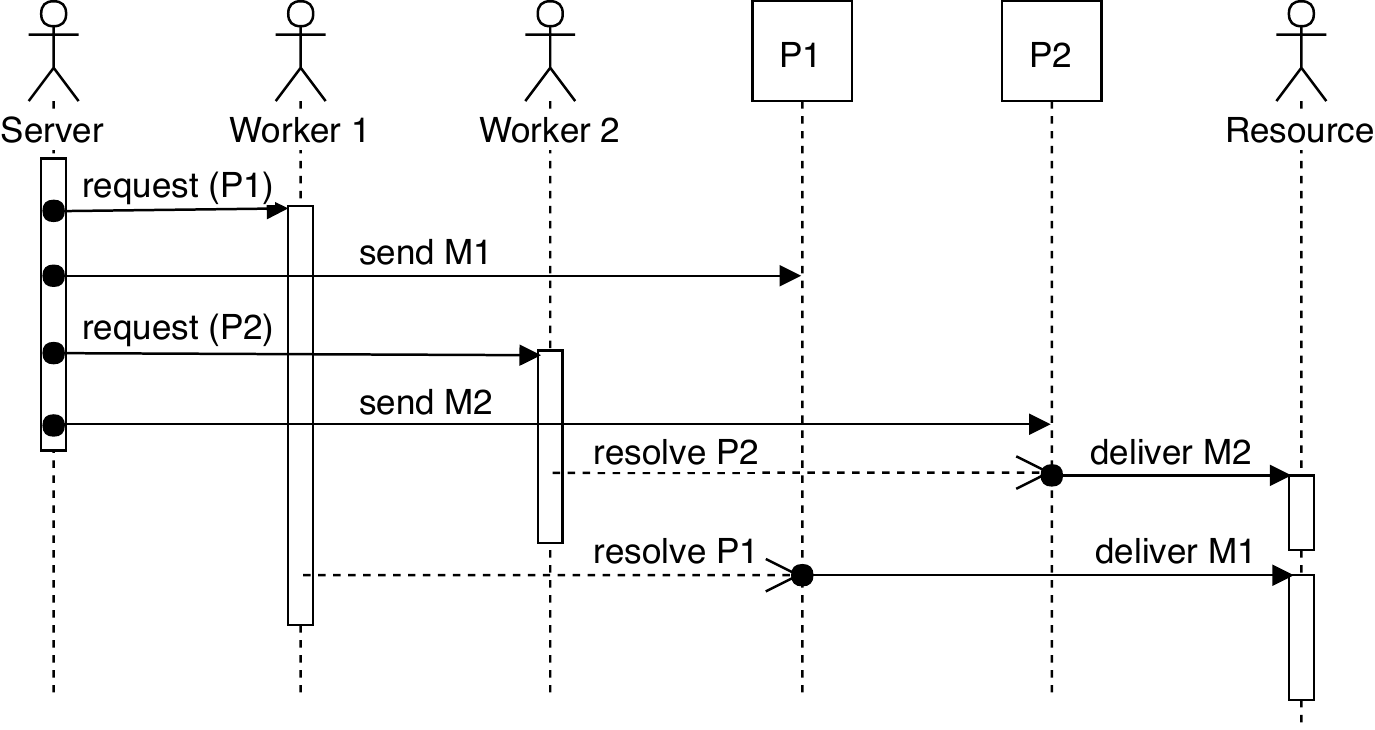}
		\caption{Promise issue with the MPL* approach, message \code{M2} is able to overtake message \code{M1}. 
		}
		\label{fig:promiseproblem}
\end{figure}

\Cref{fig:promiseproblem} gives an example of a scenario where replay
using only message sender information would not suffice for an actor-based language,
because it is not eliminating all non-determinism.
	The \code{Sever} actor creates two promises \code{P1} and \code{P2} and then sends a \code{request} message with promise \code{P1} as an argument to the \code{Worker1} actor, and 
	a message \code{M1} to promise \code{P1}. 
	This is repeated with \code{Worker2}, \code{P2} and \code{M2}.
	In our example, 
	\code{Worker2} resolves \code{P2} to \code{Resource}, 
	causing the message \code{M2} stored in \code{P2} to be delivered to it. 
	Later, \code{Worker 1} also resolves its promise (\code{P1}) to the same \code{Resource}, and 
	message \code{M1} is delivered. 
	Despite being sent first, 
	\code{M1} is processed after \code{M2}, 
	as in our scenario the processing order depends on which promise is resolved first. 
	This makes the message ordering non-deterministic when there is a race on which promise is resolved first. 
	
	In short, MPL* and Actorverse cannot reliably replay a program similar to the scenario of \cref{fig:promiseproblem}, 
	as \code{Server} is the sender of both messages, and 
	replay cannot distinguish between \code{M1} and \code{M2}.

\paragraph{Issue 2: Recording non-deterministic input.}
As stated before, pure ordering-based replay cannot deal with non-determinism caused by external inputs. 
Ordering-based replay variants devised for MPI programs can deal with one source of non-determinism: messages exchanged between processes.
In particular, MPL* does not trace non-deterministic contents of messages and as such, 
it does not support replay of I/O operations. 

On the other hand, content-based replay variants devised for JavaScript's event loop concurrency can deal with non-determinism caused by external input. 
Jardis\citep{barr:2016:time} is able to trace systems calls and I/O. 
Dolos\citep{Burg:2013:IRW} captures all I/O, user input, and other sources of non-determinism,
 such as timers for JavaScript programs.
However, both Jardis and Dolos only support a single event loop. 

It is thus an open issue to support both types of non-determinism for actor-based concurrency: message non-determinism (MPL*) and non-deterministic interactions within a turn (Jardis, Dolos).

\paragraph{Issue 3: Scale.}
With content-based replay, 
the trace contains enough information to make replay of individual actors in isolation possible.
This can be useful when the origin of a bug has been narrowed down to a few actors, 
the behavior of which can then be examined in detail without being distracted by the rest of the system. 
However, the set of problematic actors is usually unknown beforehand, rendering the approach often impractical, as it does not offer deterministic replay of all the actors in a system.
We also expect high overhead for content-based replay both in execution time and memory footprint since more events need to be recorded, for example, messages exchanged between actors.

Ordering-based replay approaches proposed in the context of message passing libraries (MPL)
seem better suited for actors.
To the best of our knowledge,
there is no existing performance comparison between the two flavors,
MPL* and MPL-ROLT.
However, MPL-ROLT suffers from scalability issues when applied to large-scale systems, 
since it needs to update the clock of a message sender, 
when the receiver's clock is greater.

This back-propagation of clocks works in the context of MPI, 
where mandatory ACK can be used.
Also, the sender requires synchronization of its mailbox to avoid clock 
updates from received messages while waiting for the ACK response.
Blocking the mailbox while sending a message may be problematic given the larger number of actors and messages found in actor programs.

Even though MPL replay approaches provide a starting point for replaying actor-based concurrent programs, they assume a coarse-grained granularity of processes and sparse use of message-based communication. 
In contrast, actors are very lightweight and are commonly used on a very fine-grained level, comparable to objects.
As such, a large number of actors can be created per VM.
Not only does this imply that the traffic generated by messages is higher than in MPL programs, 
but also that tracing needs to be optimized for events such as actor creation, messages, and I/O.

\section{Deterministic Replay for Actors}
\label{sec:solution}

The following sections present our solution to the non-determinism of high-level messaging abstractions,
input from external sources, and the scale and granularity of actor systems.

The effects of high-level messaging abstractions, such as promises, are replayed by recording and using additional information, which is discussed in the remainder of this section.

To handle non-deterministic input, we propose a design that distinguishes between synchronous and asynchronous inputs to fit well with the actor model (\cref{sec:nondet}).
Finally, to handle fine-grained actor systems, we use a compact trace format that can be recorded with a low run-time overhead and generates traces with manageable sizes (\cref{sec:tracing}).

\subsection{High-level Architecture}
\label{sec:high-level-arch}

To achieve deterministic replay,
we record the necessary information to replicate
the message execution order of an execution precisely.
To this end, we record actor creation, the message processing order,
and external non-determinism, \ie, input data.

As mentioned previously,
there is a wide range of different actor systems\citep{DeKoster:2016:YAT}.
However, some actor systems use similar implementation strategies to gain efficiency.
While they are not a precondition for our approach,
they can influence the efficiency of tracing.
One common optimization used by many actor runtimes
is that actors are scheduled on threads arbitrarily,
possibly using a thread pool.
This means actors are not bound to a specific thread.

Another common optimization is that message are processed in batches to avoid
making the actor mailbox a synchronization bottleneck.
Thus, a thread that executes the actor can take the actor's mailbox,
replace it with an empty one, and then starts executing the messages in the mailbox without having to synchronize again.

In \cref{sec:tracing}, we utilize this property to avoid redundancy in subtraces
that correspond to a batch of messages.

To minimize the perturbation introduced by tracing,
we decouple the event recording from the writing to a file.
While it is possible to store data actor-local, doing so causes memory overhead to scale with the number of actors, which is problematic for fine-grained actor-based concurrency.
Consequently, each thread that executes actors uses thread-local buffers
to store the recorded events.
One buffer records the generic events.
The other buffer records external data.
When the buffers are full, they are handed to a thread
that writes them to a file (cf. \cref{sec:buffer-management}).
The recording itself is also optimized as discussed in \cref{sec:tracing} and \cref{sec:record-ids}.
The resulting trace file can then be used
to replay the whole execution within a new process.

\subsection{Identifying Actors}

\begin{figure}
	\centering
	\includegraphics[width=0.8\linewidth]{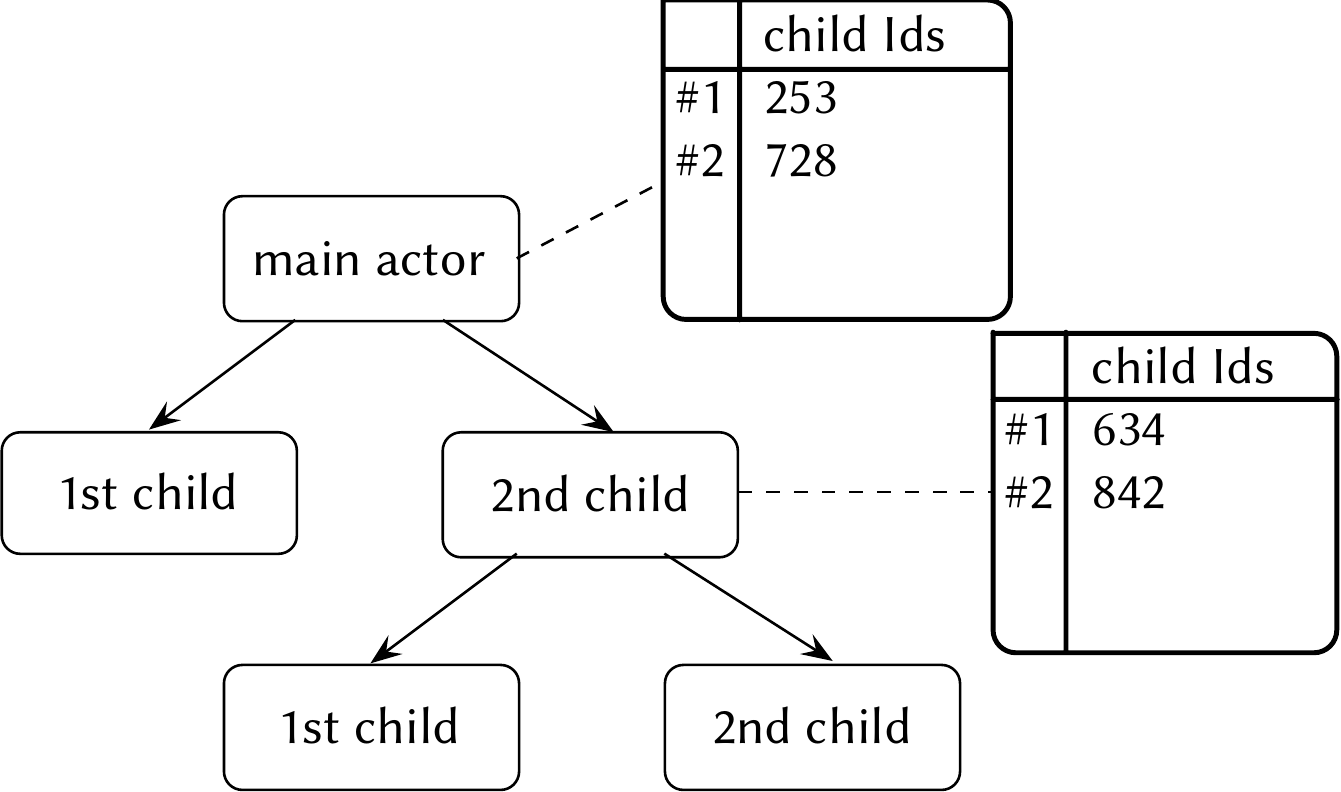} 
	\caption{Actor family tree in replay.
	Actors know which Ids to assign a new child actor. 
	}
	\label{fig:actortree}
\end{figure}

For recorded events, we need to know on which actor they happened.
For this purpose, each actor is assigned a unique integer Id (\code{ActorId}). 
To correctly assign traced data to actors during replay,
our technique has to reproduce the assignment of actor Ids.
To this end, we consider the actors that an actor spawned to be its children.
We record actor creation in our trace,
so that we can determine the Ids of an actor's children.
Using the creation order, we can reassign Ids correctly in replay.

The main actor, which is created when the program starts,
is always assigned the same Id.
We can therefore identify it and use it as a basis for identifying all its child actors.
For each actor, we keep track of how many children it created so far.
When a new actor is created during replay,
we use the actor family tree shown in \cref{fig:actortree} to look up the Id that has to be assigned to the new actor.

\subsection{Messages \& Promise Messages}
\label{sec:promises}
	For replaying normal messages, we have to record the Ids of their senders just as MPL* does.
	However,
	as shown by \cref{fig:promiseproblem} and 
	discussed in issue 1 of \cref{sub:problem},
	this is insufficient to replay high-level messaging abstractions such as promises.
	We solve the issue by recording the actor that resolved the promise,
  \ie, caused a so-called promise message to be delivered.
	
	With this additional information we are able to distinguish messages that would otherwise appear identical, as for instance in a MPL* replay. 
	In the example of \cref{fig:promiseproblem},
  we now know which worker is responsible for which message,
  and can therefore ensure that they are processed
  in the same order as in the original execution.

\subsection{Replay}
\label{sec:replay}

	\begin{figure}
		\includegraphics[ width = 0.4\textwidth]{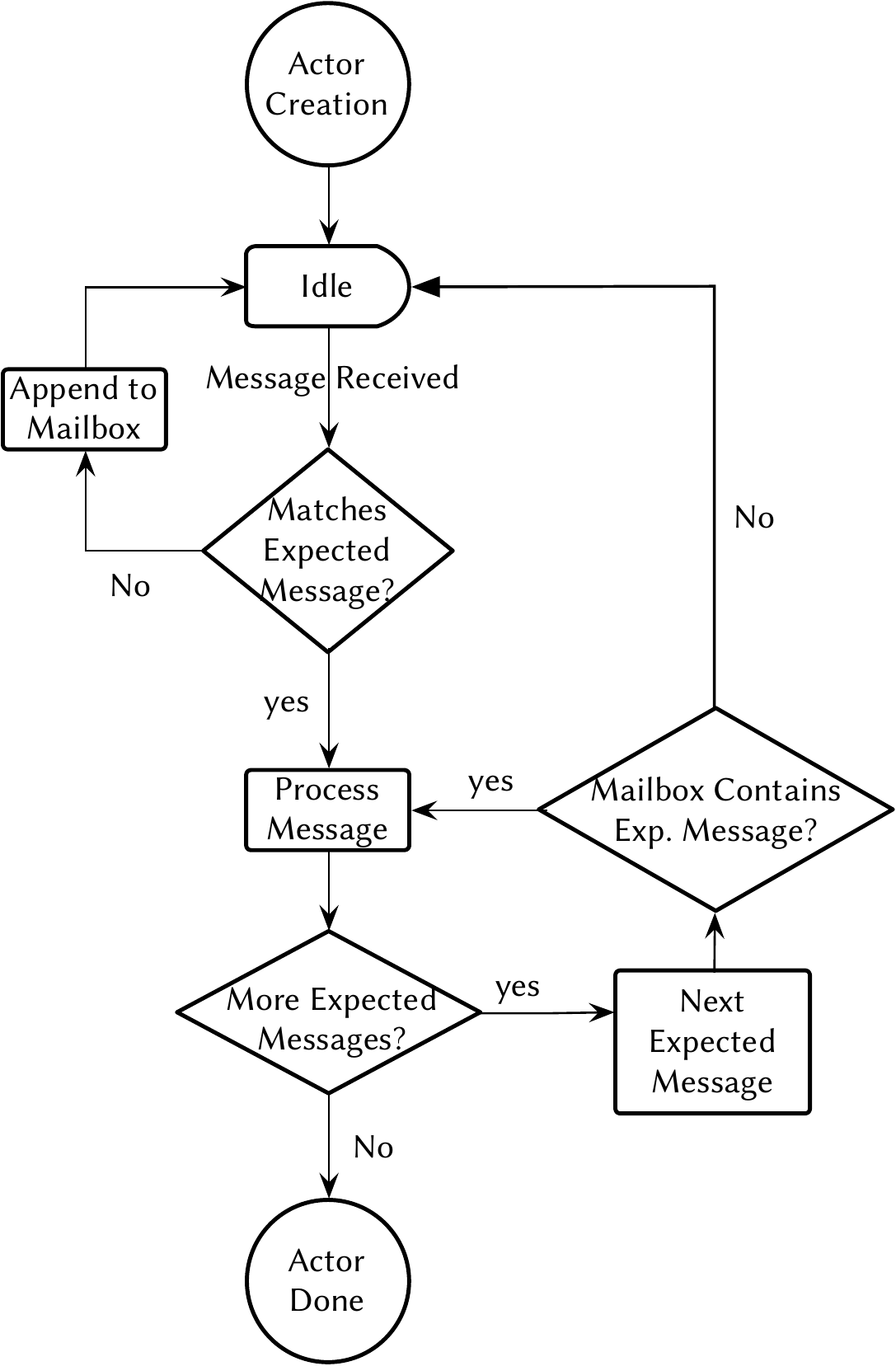} 
		\caption{Behavior of actors during replay. To reproduce the message order, actors check if the message type, sender, and (for promise messages) resolver of a message matches the expected message. Only a matching message is executed, and mismatches delayed until they match.	
	}
		\label{fig:replay-flow}
	\end{figure}

  When a program is started in replay mode,
  the interpreter loads the trace file and starts executing the program.
  Instead of relying on the normal actor implementation,
  it uses an implementation specifically adapted to replay the
  trace exactly.

  During replay, each actor holds a queue of recorded information 
  that represents the message order to reproduce.
	We call the head of this queue the \emph{expected message}.
  The expected message is either a normal message or a promise message.
  To be processed, a received message needs to match this type.
	For normal messages,
  the received message also needs to have the same sender Id. 
	Similarly, 
	for a promise message the received message needs to have
  the same sender and resolver Ids as the expected message.

	\Cref{fig:replay-flow} shows how actors behave in replay executions.
	The way an actor handles an incoming message
	depends on whether it is currently idle or processing a message.
	An idle actor will check if the received message has the sender
  and possibly resolver Id of the expected message.
	If it does, the new message will be processed right away.
	Otherwise, the message is appended to the mailbox.
	When an actor is busy and receives a message,
	the message is simply appended to the mailbox.
	When a busy actor finishes processing a message,
	it will peek at the next expected message in the queue,
	and then iterate through the mailbox in search for a matching message.
	If a match is found, the message is processed,
	otherwise the actor becomes idle and stops processing messages until a matching message is received.

\section{Capturing External Non-determinism}
\label{sec:nondet}

	Most programs interact with their environment,
	the effects of which can be non-deterministic. 
	For instance, in an HTTP server that receives requests and reacts to them,
	the request order determines the program behavior.
	Another example for external non-determinism are system calls to get the current time.
	Hence, capturing such inputs is essential
    for deterministic replay.
 
	We distinguish two ways
      non-determinsim is introduced by such interactions:
  system calls and asynchronous data sources.
  System calls are interactions with the environment
	that directly return a result, 
	such as getting the current system time, 
	or checking whether a file exists.
  Asynchronous data sources are more complex
  and introduce non-determinism
  through an arbitrary number of messages 
  that are pushed as result of a non-deterministic event.
	For example, an incoming HTTP request can cause a message
  to be sent to an actor.
	
	During recording, 
	all interactions with the environment are performed and the data needed to return results or send messages is recorded. 
	Each operation's data is assigned an Id that is used to reference it,
	and is written to a data file. 
	
	To enable tracing with minimal run-time overhead and storage use,
  we leave the decision what and how to record to the implementers of data sources.
	Hence, the tracing mechanism for external data is general enough to be used
  for a wide range of use cases.

	\subsection{System Calls}
		\label{lab:syscall}

		The system call approach targets synchronous interactions
        with non-deterministic results, 
		which are recorded.
		All system calls are expected to be implemented
    as basic operations in the interpreter
    and are executed synchronously without sending a message.
		This means that they happen as part of a turn.
		
		Each system call needs to be carefully considered for tracing,
		to prevent external data from leaking into the program uncontrolled.
		Critical objects on which the system calls operate (\eg a file handle)
        need to be wrapped, and have to be completely opaque.
        Otherwise the program can access external data that is not replayed.
		This means that all operations that involve the wrapped object are either system calls or only access fields of the wrapper.

		As a result of the tracing,
    we get an ordered sequence of system calls for each actor
    as well as the data that came from each of these calls.
		By reproducing the order of events for an actor,
		we also reproduce the order of performed system calls. 
		Hence, the result of the n-th system call by an actor is referenced by the n-th system call event in the trace.
		
		When an actor performs a system call in replay, 
		the \code{DataId} of the queues head is used to get the recorded data.
		The system call then processes that data, 
		instead of interacting with the environment,
		and thus returns the same result as in the original execution.

		The implementation of system calls is straightforward. 
		\Cref{fig:syscall} is a simple example for a system call that checks whether a path represented by a string exists in the file system.
		In the Java implementation of the system call, we insert two if clauses, 
		the first one (lines 2-4) is placed before the existence is actually checked. 
		During replay, it will get the result from the original execution and return it immediately, 
		bypassing the rest of the method. 
		The existence check is performed in line 6 and the result is stored in a variable \code{result}. 
		Finally, the second if clause (lines 6-8) is responsible for recording the result when tracing is enabled. 
		The infrastructure adds a system call event to the trace and records the result in a separate data trace.
		
		Our design focuses on the reproduction of the returned result, 
		but it is general enough to allow reproduction of other effects a system call may have on the program. 
		For instance, a system call that also resolves a promise in addition to returning a value. 
		In this case the recorded data has to contain both the result and the value used for promise resolution.
		
		\lstdefinestyle{mystyle}{
			backgroundcolor=\color{white},   
			commentstyle=\color{green},
			keywordstyle=\color{blue},
			numberstyle=\tiny\color{gray},
			stringstyle=\color{orange},
			basicstyle=\footnotesize,
			breakatwhitespace=false,         
			breaklines=true,                 
			captionpos=b,                    
			keepspaces=true,                 
			numbers=left,                    
			numbersep=5pt,                  
			showspaces=false,                
			showstringspaces=false,
			showtabs=false,                  
			tabsize=2
		}
		
		\lstset{style=mystyle}
		
\begin{figure}
\begin{lstlisting}[language=Java]
public boolean pathExists(String path) {
  if (REPLAY) {
    return getSystemcallBoolean();
  }
  boolean result = Files.exists(path);
  if (TRACING) {
    recordSystemCallBoolean(result);
  }
  return result;
}
\end{lstlisting}
\caption{Simplified example for the implementation of a path exists system call.}
\label{fig:syscall}
\end{figure}

\subsection{Asynchronous Data Source}
\label{sec:data-source}
		Input data that is not handled with system calls is generally considered
    to come from some asynchronous data source.
    In an actor system, this means the external data source is typically
    represented by an actor itself and data is propagated in the system by
		sending messages or resolving promises.
    Thus, an actor wraps the data source and makes it available to other actors via messages.

    Through this wrapping, the deterministic replay can rely in part
    on the mechanisms for handling messages and promises.
    However, we need to augment them to record the data from the external source
    when it becomes accessible to the application.
		These messages and promise messages that are sent as result of external events are marked as \emph{external messages}.
    For example, a message sent to an actor
    that is triggered by an incoming HTTP request
    will be marked as external and will contain the data of the request.
    In the trace, these messages are marked as external as well and contain
    the data Id to identify the recorded data during replay.
    They also contain a marker to identify the type of event for a data source.
    This is necessary because each data source may have multiple events of different kinds.
    The data itself is stored in a file separate from the traced events
    (cf. \cref{sec:external-data,sec:tracing}).
    
    When an actor expects an external message during replay,
    it will not wait for a message,
    but instead simulate the external data source.
    Thus, it reads the recorded data associated with the sending actor
    and the data Id in the trace.
    With this information, the replay can resolve promises and send messages
    with the same arguments as during recording.

\subsection{Combining Asynchronous Data Sources and System Calls
            to Record Used Data Only}
\label{sec:opt-data-recording}

\begin{figure}
			\centering
			\begin{subfigure}{0.8\linewidth}
				\centering
				\includegraphics[width=\textwidth]{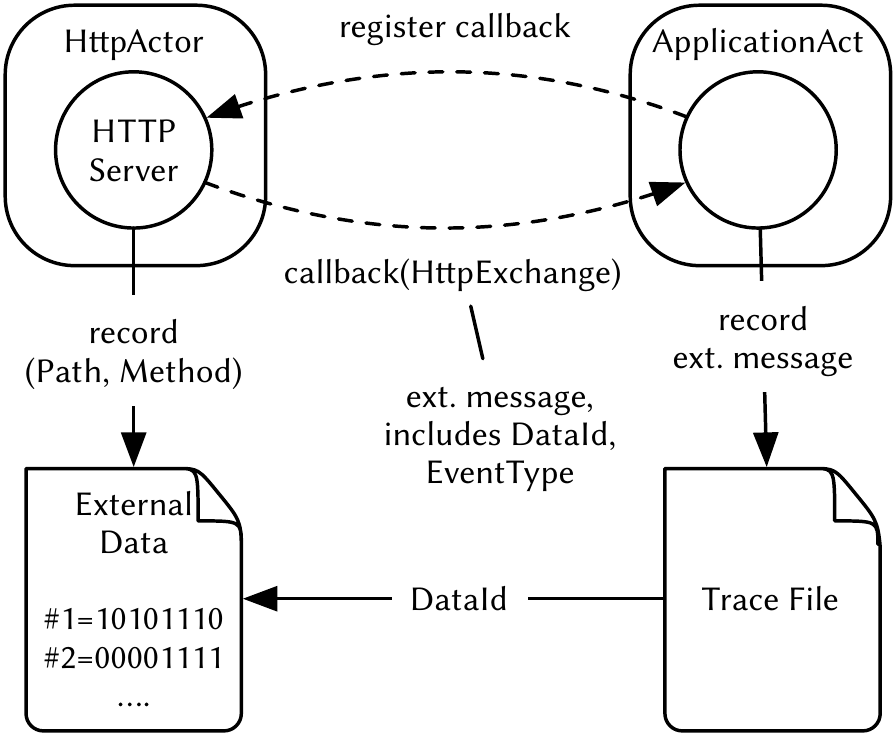}
				\subcaption{Tracing}
				\label{fig:ExtTrace}
			\end{subfigure}
			\hfill
			\begin{subfigure}{0.8\linewidth}
				\centering
				\includegraphics[width=\textwidth]{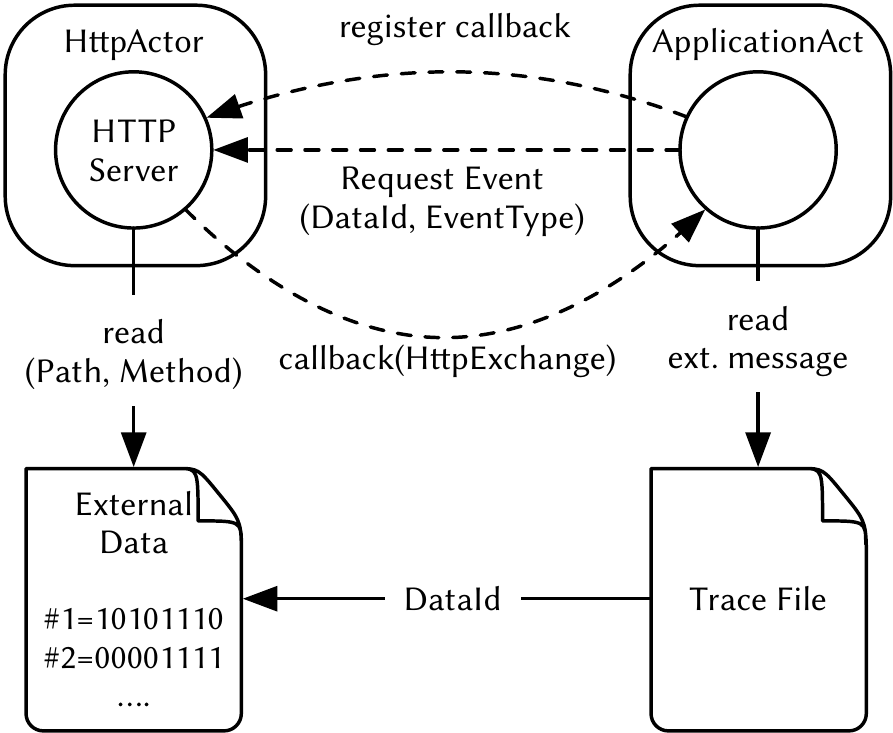}
				\subcaption{Replay}
				\label{fig:ExtReplay}
			\end{subfigure}
			
			\caption{Data flows of an HTTP server during tracing and replay. 
			Information about an incoming request is recorded in the trace, 
			this event is reproduced in replay on request of the ApplicationActor.
			}
			\label{fig:exttrace}
		\end{figure}

  Depending on the application that is to be recorded \& replayed,
  it can be beneficial to avoid recording all external data
  and instead only record the data that influenced the application,
  \ie, was used by it.
  To this end, we can combine our notion of asynchronous data sources and
  system calls.
  We detail this idea using our example of an HTTP server,
  where an application might only inspect the headers sent by a client,
  but might not need the whole body of the request.
  \Cref{fig:exttrace} gives an overview of how the system is structured to
  deal with such a scenario.

    The HTTP server is considered an external data source and is thus represented
    by its own actor.
    Application actors can register to handle incoming HTTP requests
    on certain request paths, which is a pattern common to many web frameworks.
    The server handles the incoming HTTP requests, 
    and then delegates them to the registered actor by sending a message.
    
    A request itself can be modeled as an object,
    with which an application can interact, for instance, to read the header
    or to respond with a reply to the HTTP client.
    
    To minimize the data that needs to be recorded,
    we model our data source for the HTTP server
    so that it creates a HTTP request object only with the minimal amount
    of data.
    The HTTP headers and body are only going to be recorded when they
    are accessed.
    Therefore, during recording, the initial incoming HTTP request only leads
    to the recording of the kind of HTTP request that was made,
    \eg, a get or post request, and the request path.
    This information is needed to identify the callback handlers
    that an application actor registered on the HTTP data source.
    
    As detailed in \cref{sec:data-source},
    triggering the callback handler is done via an external message.
    Thus, the recorded message contains the \code{DataId},
    which references the kind of HTTP request made and its path.
		During replay, when the HTTP server is created, 
		its actor has the same \code{ActorId} as in the recorded execution, 
		and the same callbacks are registered by the same actors in the same order.
	When the application actor expects to receive the external message,
    it looks up the data source (HTTP server)
    based on the sender's \code{ActorId},
    and requests the simulated event.
    Thus, the data source recreates the HTTP request object based on the
    \code{DataId}.
    When an application accesses for instance
    the HTTP headers and body at a later point,
    we handle these as system calls.
    Thus, during recording the header data is written
    into the trace, and
    read from the trace file during replay.

	\subsection{Format for External Data}
\label{sec:external-data}

		External data that is recorded for external events and system calls
		is stored in a separate trace file using a binary format.
		The file has a simple structure of consecutive entries with variable length.
		Each entry starts with a 4-byte \code{ActorId} for the origin of the entry.
    It is followed by the 4-byte \code{DataId},
    which is referenced by the trace entries
    for external messages and system calls.
    The length of the payload is encoded also with 4-byte field.
		The combination of \code{ActorId} and \code{DataId}
    allows it to identify a specific entry globally.

\section{Compact Tracing} 
\label{sec:tracing}

To encode trace events,
we use a binary format that can be recorded
without introducing prohibitive run-time overhead.
As mentioned in \cref{sec:high-level-arch},
we also need to account for actors being executed on different threads over time.
Both aspects are detailed below.
	
\subsection{Subtraces}

Since actors can be scheduled on different threads over time,
and we use thread-local buffers to record events,
we need to keep track of the actor that performed the events.
To avoid having to record the actor for each event,
we start a new \emph{substrace} when an actor starts executing on a thread.
Similarly, when a buffer becomes full, a new subtrace is started.

To minimize run-time overhead,
we use thread-local buffers that are swapped only when they are full.
This however means that an actor could execute on one thread,
and then on another,
and the buffer of the second thread
could be written to the file before the first one.
Thus, we need to explicitly keep track of an ordering of subtraces.
For this reason, actors maintain a counter for the subtraces.
We record it as a 2-byte Id as part of the start of subtraces.
	For well-behaved actor programs, the buffers are written in regular intervals
  and 2-byte Ids provide sufficient safety even with overflows to restore the
  original order.

\subsection{Trace Format}
\label{sec:traceEntryEnc}

Our compact binary trace format uses a one-byte header
to encode the details of a trace entry,
and then encodes entry-specific fields.
		The bits in the \code{E} event header encode
    the type of the entry,
    whether a message is marked as external,
    and the number of bytes that are used for Ids.
    \Cref{Fig:TraceFormat} visualizes the encoding.

    By encoding the Ids with flexible length,
    we can reduce the trace size significantly
    (cf. \cref{sec:record-ids,sec:evaluation}).
    Ideally, it means that an Id smaller than 256
    can be encoded in a single byte, one smaller 65536 in two bytes, and so on.

As discussed in the previous section,
we need to record the start of subtraces, actor creation, messages,
promise messages, and system calls.
Their specific fields are as follows:

		\begin{figure}
			\centering
			\includegraphics[width=0.98\linewidth]{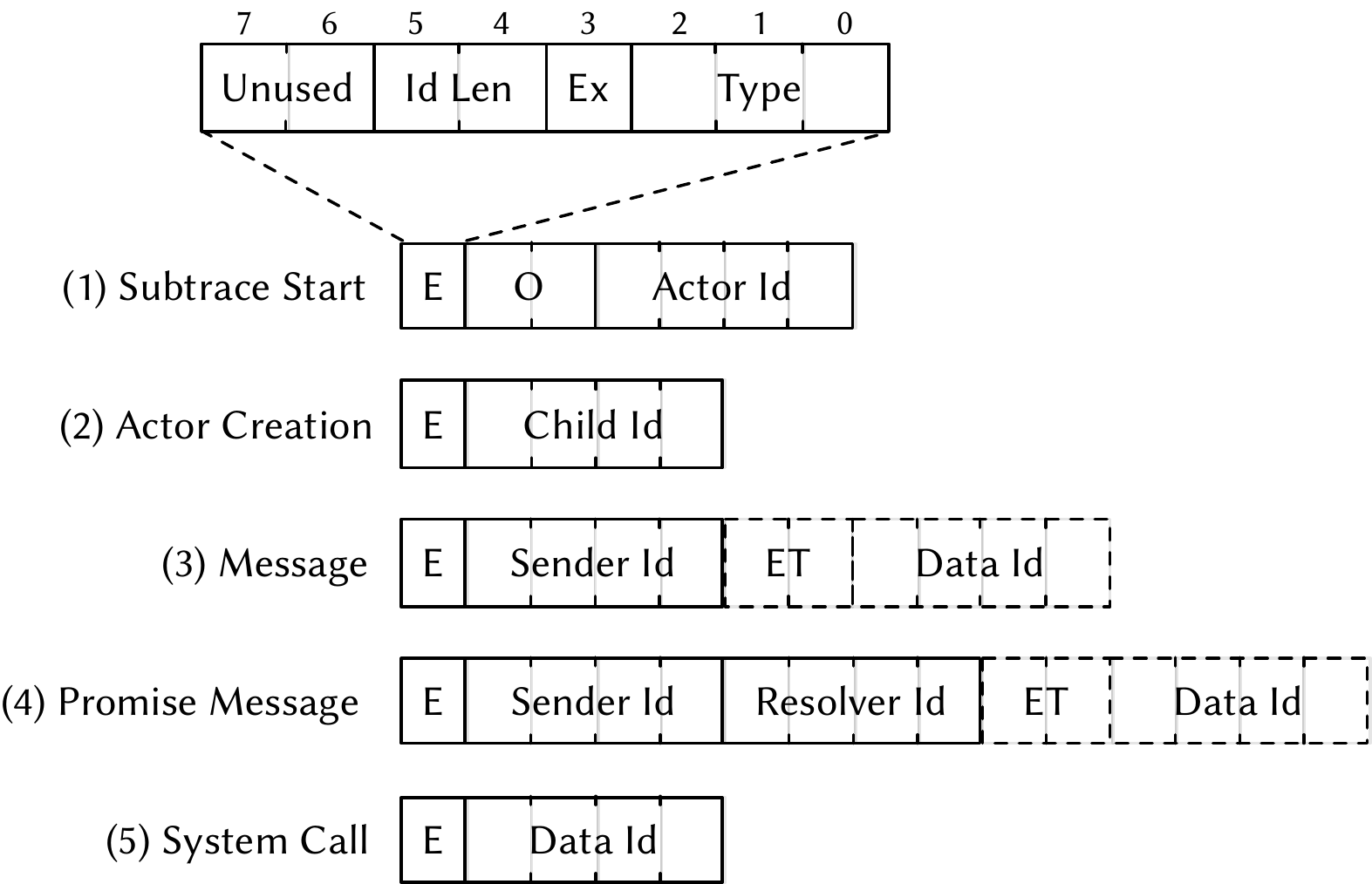}
			\caption{Sketch of the encoding of trace entries with 4-byte Ids.
              The \code{EventType} and \code{DataId} fields of messages
              and promise messages are only needed
              when they are marked as external in the header.}
			\label{Fig:TraceFormat}
		\end{figure}

		\paragraph{(1) Subtrace Start} A subtrace start indicates the beginning
      of each subtrace to associate all events
      within it with the given \code{ActorId}.
      \code{O} is the 2-byte ordering Id
      to restore the correct order of subtraces before replaying them.
		
		\paragraph{(2) Actor Creation}
      The actor creation entries correspond to when a child actor is spawned.
      It includes the Id of the new actor (\code{Child Id})
      so that we can construct the parent-child tree of actors for replay and
      reassign Ids to each actor.

		\paragraph{(3) Message \& (4) Promise Message}
		\label{lab:ExternalMessage}
		Message entries correspond to the messages processed by the actor
    of a subtrace. 
		The \code{SenderId} identifies the actor that sent the message. 
		Promise messages also include the \code{ResolverId}
    to identify the actor that resolved the promise. 
		
		External messages are marked by the \code{Ex} bit in the event header
    and record \code{EventType} (ET in \cref{Fig:TraceFormat})
    and \code{DataId}.
		The \code{EventType} identifies the kind of external event,
    \eg, an HTTP request.
    It is used to distinguish different kind of events from the same source.
    The 4-byte \code{DataId} references the data for the external event.

		\paragraph{(5) System Call}
		System call entries record the \code{DataId} to identify the data.
    Note that the order of trace entries is in most cases sufficient
    to recreate a mapping during replay.
    Identifiers are only introduced for cases where the ordering is insufficient.

\section{Implementation}

We implemented our record \& replay solution for
communicating event-loop actors in \SOMns.
\SOMns is written in Java as a self-optimizing abstract syntax tree (AST) interpreter\citep{Wurthinger:2012:SelfOptAST}
using the Truffle framework and Graal just-in-time compiler\citep{Wurthinger:2017:PPE}.
This allows us to integrate record \& replay
directly into the language implementation.
The tracing is added as nodes that specialize themselves (\ie optimize) based on the inputs they encounter.
This means, the tracing is compiled together with the application code and
executes as highly optimized native code,
which reduces run-time overhead.
Our implementation optimizes recording of Ids
and delegates the writing of trace data to a background thread,
which we detail below.

\subsection{Optimized Recording of Ids}
\label{sec:record-ids}

As seen in \cref{sec:traceEntryEnc},
identifiers (Ids) are the main payload for trace entries.
Thus, efficient recording of Ids is crucial for performance.
To minimize the trace size,
we decided to encode them in smaller sizes if possible.
However, in a naive implementation this would increase the run-time overhead
significantly, because for each Id we would need to check how to encode it resulting in complex control flow
possibly limiting compiler optimizations.

With the use of self-optimizing nodes,
we can avoid much of the complexity of writing Ids.
A program location that for instance spawns an actor can thus specialize
to the value range of Ids it has seen.
To minimize the overhead,
a node specializes to the value range that fits all previously seen Ids.
Thus, if only an Id 34 has been recorded,
the node specializes to check that the Id matches the 1-byte Id range
and to write it.
If Ids 34 and 100,000 has been seen,
the node specializes to check that the Id can be stored in 3 bytes
and writes it.
In case an Id is encountered that does not fit into the given number of bytes,
the node replaces itself with a version that can write longer Ids.
This will also invalidate the compiled code,
and eventually result in optimized code being compiled.

While this approach does not achieve the smallest possible trace size,
it reduces the run-time overhead.
We evaluate the effectiveness of our optimization and its effect on the performance in \cref{sec:evaluation}.

\subsection{Buffer Management}
\label{sec:buffer-management}
For our tracing of regular events,
we use the following thread-local buffer approach as described by \citet{Lengauer:2015:AEO:2668930.2688037}.
By using thread-local buffers, we avoid synchronization for every traced event. 
Buffers that are not currently used by a thread are stored in two queues, one containing full, 
and the other containing empty buffers. 
When a thread's trace buffer does not have enough space for another entry, 
it is appended to the full queue, 
and the thread takes its new buffer from the empty queue. 
The full queue is processed by a background thread
that writes the trace to a file.

For external data, we use separate buffers and a separate queue.
As external data can be of any size, we allocate buffers on demand
and discard them when they are no longer needed.

The \emph{writer thread} that persists the trace
also processes the queue for external data.
Once a buffer is written, it is added to the queue of empty buffers for
trace data or discarded for external data.

To avoid slowing down application threads with serialization
and conversion operations,
they are done by the writer thread.
The application threads hand over the data without copying
whenever it is safe to do so.
For instance, for our HTTP server data source, this is possible 
because most data is represented as immutable strings.
Data sources that use complex objects use serializers
that are handed over to the writer thread together with the data.
This makes it possible to persist also complex data on the writer thread.

\section{Evaluation}
\label{sec:evaluation}

This section evaluates the run-time performance and trace sizes
of our implementation in \SOMns using the Savina benchmark suite
for actors\citep{Imam:2014:SAB}, 
and Acme-Air as an example for a web application. 
We also use the \emph{Are We Fast Yet} benchmarks
to provide a baseline for the \SOMns performance\citep{marr2016cross}.

\subsection{Methodology}

As \SOMns uses dynamic compilation,
we need to account for the VM's warmup behavior\citep{Barrett:2017:VMW}.
The Savina and Are We Fast Yet benchmarks run for 1000 iterations within the
same VM process using ReBench\citep{ReBench:2018}.
Since we are interested in the peak-performance
of an application with longer run times, we discount warmup.
We do this by inspecting the run-time plots for all benchmarks,
which indicates that the benchmarks stabilize after 100 iterations. 

For Acme-Air, we use JMeter\citep{halili2008apache} to produce a predefined workload of HTTP requests.
The workload was defined by the Node.js version of Acme-Air.
JMeter is configured to use two threads to
send a mix of ca. 42 million randomly generated requests
based on the predefined workload pattern.
After inspecting the latency plots, we discarded the first 250,000 requests
to exclude warmup.

The Savina and Are We Fast Yet benchmarks were executed on a machine with
two quad-core Intel Xeons E5520, 2.26 GHz with 8 GB RAM,
Ubuntu Linux with kernel 4.4, and Java 8.171.
Acme-Air was executed on a machine with a four-core Intel Core i7-4770HQ CPU,
2.2 GHz, with 16 GB RAM, a 256 GB SSD, macOS High Sierra (10.13.3),
and Java 8.161.
In both cases, we used Graal version 0.41.

In \cref{sec:requirements}, we defined the performance goals
of a tracing run-time overhead of less than 25\%
for message intensive programs,
\ie, microbenchmarks such as from the Savina benchmark suite.
Savina falls into this category
as many of the benchmarks perform little computation,
for instance, in the counting benchmark
200,000 messages are sent to an actor who increments a counter.
Furthermore, we aimed for a tracing mechanism that produces under 250 MB/s
of trace data to be practical on today's developer machines.
For larger systems, which are not dominated by message sends,
we aim for run-time overhead that is in the range of 0-3\%.

To assess whether we reach these goals,
we measure the overhead of tracing on the benchmarks while restricting
the actor system to use a single thread.
This is necessary to measure the actual tracing overhead.
Since some benchmarks are highly concurrent,
running on multiple threads can give misleading results.
One issue is that some of these benchmarks have very high contention
and any overhead in the sequential execution can result in a speedup
in the parallel execution,
because it reduces contention and the number of retries of failed attempts.

\subsection{Baseline Performance of \SOMns}

\begin{figure*}
	\begin{minipage}{.2\textwidth}
		\centering
		\AwfyBaseline{}
		\captionof{figure}{Performance comparison with other languages. \SOMns performs similar to Node.js.}
		\label{fig:awfy-baseline}
	\end{minipage}
	\hspace{5mm}
	\begin{minipage}{.7\textwidth}
		\centering
		\SavinaBaseline{}
		\captionof{figure}{Performance of Savina benchmarks in different actor languages for different numbers of Cores.}
		\label{fig:savina-baseline}
	\end{minipage}
\end{figure*}

To show that \SOMns reaches a competitive baseline performance,
and is a solid foundation for our research,
we first compared it to Java, Node.js, and Scala.

The sequential performance of \SOMns, 
as measured with the Are We Fast Yet benchmarks,
is shown in \cref{fig:awfy-baseline}.
While \SOMns is not as fast as Java, 
it reaches the same level of performance as Node.js,
which is used in production environments.
This indicates that the sequential baseline performance
of \SOMns is competitive with similar dynamic languages.

To ensure that \SOMns' actors are suitable for this work,
we compare its actor performance with other actor implementations,
based on the Savina benchmark suite. 

Unfortunately, the benchmarks are designed 
for impure actor systems, such as Akka,
Jetlang, and Scalaz.
This means, some of the benchmarks rely on shared memory.
Thus, we had to restrict our experiments 
to a subset of 18 benchmarks from the total of 28,
as the other ones could not be ported to \SOMns,
because it does not support shared memory between actors.

The results of our experiments with Savina benchmarks are
shown in \cref{fig:savina-baseline}
and indicate that \SOMns reaches the performance of other widely used actor systems on the JVM.
Hence, it is a suitable foundation for this research.

\subsection{Tracing Savina}

\begin{figure}[htb]
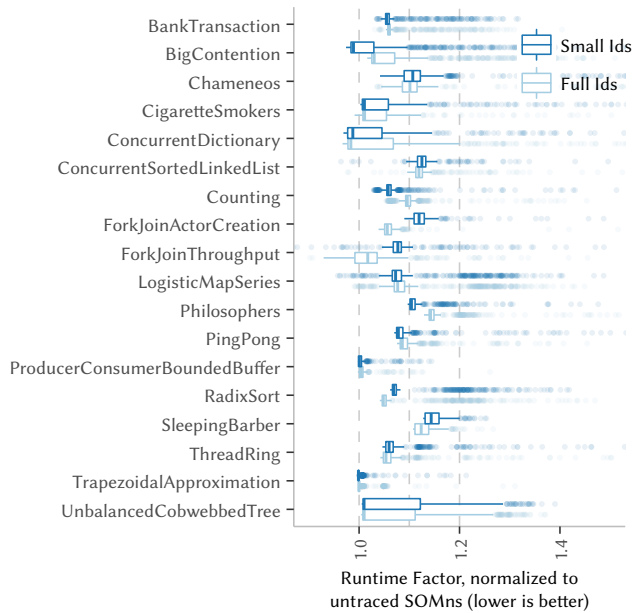

  \centering
	\SavinaTracing{}
	\caption{
    Performance of traced executions of the Savina benchmarks
    using a single thread. Results are normalized to the untraced execution.}
	\label{fig:SavinaTracing}
\end{figure}

\Cref{fig:SavinaTracing} shows the run-time overhead of tracing.
It includes the results for recording full Ids, \ie
all Ids are recorded with 4 bytes,
and the optimized version where the Ids are encoded with fewer bytes
if possible (cf. \cref{sec:record-ids}).
The average overhead for tracing with full Ids is
\TraceOverheadFidGMeanP (min. \TraceOverheadFidMinP,
max. \TraceOverheadFidMaxP).
As seen in \cref{tab:traceSize}, the benchmarks produce up to \TraceRateFidMaxMBS of data.
Applying our optimization for recording small Ids,
the average overhead for tracing is only minimally higher with \TraceOverheadSidGMeanP (min. \TraceOverheadSidMinP, max. \TraceOverheadSidMaxP).
Furthermore, the maximal data rate goes down to \TraceRateSidMaxMBS.

With these results, we fulfill our goal of having less than 25\% overhead
for programs with high message rates,
and to produce less than 250\,MB/s of trace data.

As seen from \cref{tab:traceSize},
effectiveness of using \emph{small Ids} depends on the benchmark.
\emph{TrapezoidalApproximation} for instance, has an insignificant reduction in trace size. 
Other benchmarks, such as \emph{Counting} and \emph{RadixSort}, have a near halved data-rate.
Due to the minimal performance impact,
we consider using small Ids beneficial.

\begin{table}[htb]
	\centering
  \SavinaTraceDataTable{}
	\caption{Trace production per second over 1000 benchmark iterations.}
	\label{tab:traceSize}´
\end{table}

\subsection{Tracing Acme-Air}

\begin{figure}
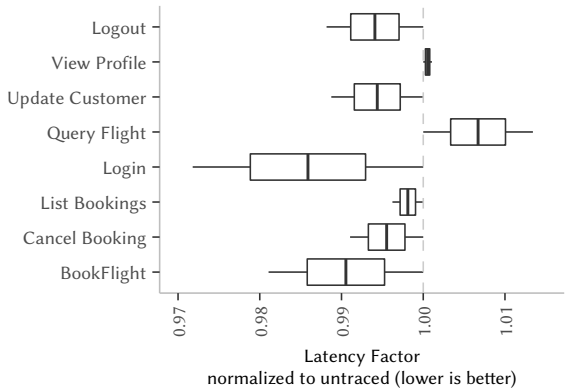

	\AcmeTracing{}
	\caption{Latency of different request types in traced executions,
           normalized to untraced executions.}
	\label{fig:acme}
\end{figure}

Acme-Air is a benchmark representing a server application implemented with micro-services\citep{Ueda:2016:AcmeAir}. 
It models the booking system of a fictional Airline.
Acme-Air is available on \citeurl{GitHub}{Acme-Air repository}{}{}{https://github.com/acmeair/} for Java and Node.js.
The JavaScript version of Acme-Air served as the basis for our \SOMns port. 
We stayed true to the original design, 
in which a single event loop is used to process requests.
Instead of using a stand-alone database, 
we used an embedded \citeurl{Derby}{Apache Derby}{}{}{https://db.apache.org/derby/index.html} database.
The database was reset and loaded with data before each benchmark to factor out its potential influence on the results.

JMeter measures the latency for each request it makes.
Since the highest resolution is 1\,ms, some results are rounded to 0\,ms.
The predefined workload uses different frequencies
for the different possible requests seen in \cref{fig:acme}.
For instance, Query Flight is the most common one representing \AcmeQueryFP of all requests.

\Cref{fig:acme} shows the effect of tracing on the latency of different requests.
Tracing \emph{small Ids} results in a maximum overhead of \AcmeLatencyTracedMaxP
(geomean. \AcmeLatencyTracedGMeanP, min. \AcmeLatencyTracedMinP).
We consider average negative overhead, \ie speedup,
an artifact of dynamic compilation.
The compiler heuristics can trigger slightly differently,
which leads to differences in the applied optimizations, native code layout,
and caching effects.

The trace size for a benchmark execution is 993\,MB in total, with an observed data-rate of 1.4\,MB/s.
External data is responsible for the majority of the trace size (88\%, 872\,MB), while tracing of internal events accounts for 121\,MB.
Variations in trace size between different benchmark executions were negligible, \ie below 1\,MB.

The Acme-Air results suggest that real-word applications tend to have lower data-rates and overhead than most Savina benchmarks. 
The maximal overhead of \AcmeLatencyTracedMaxP suggests, however, that
tracing larger application has likely minimal impact and is thus practical,
meets our goals, and is comparable to systems that are not optimized for
the fine granularity and high message rate of actors.

\section{Related Work} 

This section discusses related work in addition to the record \& replay approaches analyzed in \cref{sub:rr}.
We compare our solution to work that employs similar tracing techniques: record \& refine, back-in-time debugging, shared memory, and profiling.

\subsection{Record and Refinement}

As explained in \cref{sec:towards}, record \& replay approaches record a program trace,
which is then used during replay to guide execution and reproduce non-deterministic behavior (such as input from external sources and scheduling of messages) in a deterministic way.
Such deterministic replay can then also provide access
to more detailed information after the original execution finished.
\citet{Felgentreff:2017} define this process as \emph{record and refine}.

Record \& refine enables low-overhead
postmortem debugging.
Thus, during recording, only the minimum necessary data to reproduce the desired
parts of a program execution is recorded,
\ie, to avoid non-determinism during replay.
All additional data, for instance to aid debugging,
can be obtained during replay execution.
We apply the same idea to \SOMns.
During recording, only the minimal amount of information is retained
and during replay, all features of the \Kompos debugger are supported.

\subsection{Back-in-Time Debugging}

Unlike record \& replay, back-in-time debugging takes snapshots of the program state at certain intervals, and they offer time travel by replaying execution from the checkpoint before the target time. 

\paragraph{Jardis}
Jardis\citep{barr:2016:time} provides both time-travel debugging
and replay functionality for JavaScripts event loop concurrency. 
It combines tracing of I/O and system calls with regular heap snapshots of the event loop.
It keeps snapshots of the last few seconds, allowing Jardis to go back as far as the oldest snapshot,
and discard trace data from before that point. 
While this keeps the size of traces and snapshots small,
it limits debugging to the last seconds before a bug occurs.
This may be a problem as the distance between root cause and actual issue is typically large in concurrent programs~\citep{Perscheid:2016}. 

Jardis reports a run-time overhead of \(\leqslant\)2\%
for compressed trace sizes of below 9 MB for 2-4 second runs.
For Acme-Air, our approach has a data rate (1.4\,MB/s) lower than the one of Jardis. 
As such, our impact on the performance of the benchmark is competitive.

\paragraph{Actoverse}
Actoverse\citep{shibanai2017actoverse} also provides both time-travel debugging
and record \& replay for Akka actors.
Unlike Jardis or our solution,
Actoverse is implemented as a library and uses annotations to mark fields to be recorded.
A snapshot of those fields is saved when sending and after processing messages.
The order of messages and snapshots is determined with Lamport clocks to avoid a global clock.
While performance is not reported, the memory usage is indicated with about
5 MB for 10,000 messages. 
Our ordering-based approach requires only about 2-15 byte per message.

\paragraph{CauDEr}
CauDEr\citep{lanesecauder} is a reversible debugger for Erlang.
It is able to undo actions and step backwards in the execution
by relying on reversible execution semantics.
CauDEr currently only addresses a subset of Erlang and focuses on 
the semantic aspects of reversible execution for debugging.
Therefore, it can help in correctness considerations,
but does not focus on enabling the debugging of larger systems as our work does.

\subsection{Shared Memory}

There is a lot of work on record \& replay
  in the context of shared memory.
Generally, 
  shared memory record \& replay
  reproduces 
  the order of synchronization operations 
  and accesses to shared memory.
The used techniques are very diverse,
  as is their impacts on run-time performance,
  which can range from negligible overhead\citep{Liu:2018:iReplayer, mashtizadeh2017Castor}
  to 35x overhead\citep{leblanc:1987:debugging}.
Castor\citep{mashtizadeh2017Castor}
  can use transactional memory 
  to improve its performance.
iReplayer\citep{Liu:2018:iReplayer} 
  records explicit synchronization,
  regularly creates snapshots of program state,
  and provides in-situ replay, 
  \ie within the same process.
LEAP\citep{Huang:2010:LEAP}
  uses static analysis to determine
  what is shared between threads.
Unfortunately, static approaches can introduce synchronization
  on all operations with a shared field.
For the actor model, 
  this synchronization corresponds to one global lock for all
  mailboxes.
In the actor model
  synchronization on
  mailboxes and promises 
  are essential.
They correspond roughly to the
  tracing in \SOMns.
Hence, shared-memory approaches 
  conceptually have to record 
  at least as many events
  as our actor tracing.
However, the actor model ensures that there are no races on shared memory,
which would need to be traced.
For pure actor models as in \SOMns,
shared memory approaches are therefor likely having the same or additional overhead.
For impure actor models, it seems beneficial to find
ways to combine actor and shared memory record \& replay techniques.

\subsection{Profiling}
We now discuss related work in the context of profiling for actor-based concurrency.

\paragraph{Profiling of Akka Actors}

\citet{Rosa:2016:PAU,Rosa:2016:APV} profile the utilization and communication of Akka actors
based on platform-independent and portable metrics.
An application collects profiling information in memory.
On termination, it generates a trace file
that can be analyzed to determine performance bottlenecks.
It tracks various details including message counts and executed bytecodes.
To attribute this information precisely, it maintains a shadow stack.
In contrast to this, \SOMns records the ordering of messages, their processing,
and any external input.
Since offline analysis is not a direct goal,
\SOMns does not need a shadow stack,
but could provide such information during replay execution.
For replaying, however, we need to record the events
instead of just counting them.
Since \citet{Rosa:2016:APV} aimed for platform-independent and portable metrics,
run-time performance was not a major concern.
They observed a run-time overhead
of about \RosaGMeanX (min. \RosaMinX, max. \RosaMaxX).\footnote{Results unpublished, from private communication.}
However, these numbers include instrumentation overhead and do not directly
compare to the overhead of long-running applications, which is probably much lower.

\paragraph{Large-scale Tracing}
Lightbend  \citeurl{Telemetry}{Telemetry}{Lightbend}{2018-05-03}{https://developer.lightbend.com/docs/cinnamon/2.5.x/instrumentations/akka/akka.html}
offers a commercial tool for capturing metrics of Akka systems.
The provided actor metrics are based on counters, rates, and times.
For example, it records mailbox sizes, the processing rate of messages,
and how much time messages spent in the mailbox.
The run-time overhead can be finely adjusted by selecting the elements that are to be traced,
and possibly a sampling granularity.
This seems to be a standard approach for such systems and is also used by
tracing systems based on the \citeurl{OpenTracing standard}{OpenTracing.io}{}{2018-05-03}{http://opentracing.io/}
or Google's Dapper\citep{sigelman2010dapper}.
However, since we want to eliminate all non-determinism,
doing selective or sample-based tracing is not an option.

\section{Conclusion and Future Work}

To better handle the complexity of concurrency and 
avoid dealing with threads and locks,
developers embrace high-level concurrency models such as actors. 
Unfortunately, actor systems usually have a limited number of supporting tools, 
making them hard to debug. 
In this paper, 
we presented an efficient record \& replay approach 
for actor languages
letting the programmer debug non-deterministic concurrency issues
by replaying a recorded trace.

Our approach is able to replay 
high-level messaging abstractions
such as promises
by recording extra information.
Non-deterministic inputs 
are recorded and replayed deterministically.
In addition, our approach 
scales to a high number of actors and exchanged messages
through its low execution time overhead
and its compact trace.

We evaluated the performance of our approach with the Savina benchmark suite, 
the average tracing run-time overhead is \TraceOverheadSidGMeanP (min. \TraceOverheadSidMinP, max. \TraceOverheadSidMaxP). In the case of the modern web application Acme-Air, our approach showed a maximum increase in latency of \AcmeLatencyTracedMaxP and about 1.4\,MB/s of trace data.

\paragraph{Applicability to Actor Models}

We argue that our approach is general enough to be applied to all forms of message processing (continuous/blocking and consecutive/interleaved).
The main reason is that we record the order in which messages are processed.
Thus, our approach is independent of any variation in selecting
which message may be executed next
and all selections, blocking, or interleaving already happened.
Hence, all those mechanisms determining the message order in the original execution do not have to be reproduced, 
as we already have the final ordering,
which is replayed in a re-execution.
A requirement for our approach is, however, that
actors are isolated and shared memory is not allowed
because we do not track races on shared memory.

\paragraph{Future work: Long Running Applications}

Although our approach is able to scale up 
in term of the number of actors and exchanged messages, 
it is currently not suitable for applications 
that run for extended periods of time. 
The trace recorded by our approach keeps growing 
as the program runs, 
at some point the trace will become too large for the disk. 
Besides the problem of growing traces,
there is also the practical issue of replaying such a program.
Replay of a program that has been running for such
a long period of time will take a similar (or higher) amount of time.

One solution is to create snapshots of the programs state at regular intervals.
Each time a snapshot is created, 
previous trace data can be discarded.
Replay can then start at the last snapshot before a failure,
and allows developers to investigate the cause.
To minimize traces further, we could apply simple compression too. 

\paragraph{Future work: Replay Performance}
Currently, our replay implementation parses the entire trace on startup.
This comes with a high memory overhead, 
and causes scalability issues with the employed data structures.
Replay scalability can be improved by parsing the trace on-the-go.
By dividing the parsing effort across the replay execution,
startup time and memory overhead can be reduced.

\paragraph{Future work: Partial Replay}
Partial replay of an execution
can enable debugging and testing techniques, 
such as regression tests 
and exploration of different interleavings.
The biggest challenge for partial replay is external non-determinism 
when switching from replay to free execution.

\begin{acks}                            
  This research is funded by a collaboration grant of the Austrian Science Fund (FWF) and the Research Foundation Flanders (FWO Belgium) as project I2491-N31 and G004816N, respectively.
\end{acks}

\bibliography{paper}

\end{document}